\documentclass[aps]{revtex4}

\usepackage{amsmath,amssymb}
\usepackage{graphicx}

\newcommand{\beqn} {\begin{equation}}
\newcommand{\eqn} {\end{equation}}

\def \beq{\begin{equation}}
\def \eeq{\end{equation}}
\def \bea{\begin{eqnarray}}
\def \eea{\end{eqnarray}}

\def \bet0{\beta_0}
\def \bet1{\beta_1}
\def \simgt{\,\rlap{\lower 7.5 pt\hbox{$\mathchar \sim$}}\raise 3 pt \hbox{$>$}\,}
\def \simlt{\,\rlap{\lower 7.5 pt\hbox{$\mathchar \sim$}}\raise 3 pt \hbox{$<$}\,}
\def\lsim{\raise0.3ex\hbox{$<$\kern-0.75em\raise-1.1ex\hbox{$\sim$}}}
\def\gsim{\raise0.3ex\hbox{$>$\kern-0.75em\raise-1.1ex\hbox{$\sim$}}}

\begin{document}


\markboth{Soltz, DeTar, Karsch, Mukherjee, Vranas}{LQCD Thermodynamics with Physical Quark Masses}

\title{Lattice QCD Thermodynamics with Physical Quark Masses}

\author{R.A. Soltz$^{\rm 1}$, C. DeTar$^{\rm 2}$, F. Karsch$^{\rm 3,4}$, Swagato Mukherjee$^{\rm 3}$, and P. Vranas$^{\rm 1}$}

\affiliation{$^{\rm 1}$ Nuclear and Chemical Science Division, Lawrence Livermore National Laboratory, Livermore CA 94550, USA}
\affiliation{$^{\rm 2}$ Physics Department, University of Utah, Salt Lake City, UT 84112, USA}
\affiliation{$^{\rm 3}$ Physics Department, Brookhaven National Laboratory,Upton, NY 11973, USA}
\affiliation{$^{\rm 4}$ Fakult\"at f\"ur Physik, Universit\"at Bielefeld, D-33615 Bielefeld, Germany}

\begin{abstract}
Over the past few years new physics methods and algorithms as well as
the latest supercomputers have enabled the study of the QCD thermodynamic 
phase transition using lattice gauge theory numerical simulations with 
unprecedented control over systematic errors.  This is largely a consequence 
of the ability to perform continuum extrapolations with physical
quark masses.  Here we review recent progress in lattice QCD thermodynamics,
focussing mainly on results that benefit from the use of 
physical quark masses: the crossover temperature, the equation 
of state, and fluctuations of the quark number susceptibilities.  In addition, we place
a special emphasis on calculations that are directly relevant to the study 
of relativistic heavy ion collisions at RHIC and the LHC.
\end{abstract}


\maketitle
\tableofcontents

\section{INTRODUCTION}

Quantum Chromodynamics (QCD), the prevailing theory of the strong interaction in nuclear physics, is one of the most successful and most challenging theories in physics.  At low temperatures, chiral symmetry breaking
gives rise to most of the mass in the visible universe, and at high temperatures it predicts the existence of a plasma of quarks and gluons~\cite{Gross:1981kh,Shuryak:1980bd}, a condition achieved in the first microsecond after the big bang as well as in nuclear collisions at the Relativistic Heavy Ion Collider (RHIC) and the Large Hadron Collider (LHC).  However, calculations at finite temperatures in the vicinity of the transition between hadronic matter and quark gluon plasma require computationally challenging techniques in lattice gauge theory~\cite{Wilson:1974ji,Creutz:1980cw}.  Steady progress in this field relies upon continuing advances in both algorithms and computing.  Recent improvements in fermion actions and computing capabilities have enabled lattice gauge calculations to perform the first reliable continuum extrapolations with physical quark masses.  The QCD transition observed in heavy ion collisions is now firmly established as a crossover at zero baryon density, and several groups have produced continuum extrapolations for the crossover temperature and equation of state (EoS) with physical quark masses.  These calculations are consistent within their respective uncertainties, which now have well defined statistical and systematic contributions.  Furthermore, access to high statistics data from heavy ion collisions coupled with new analysis techniques and improvements in hydrodynamic modeling tools have greatly enhanced the connection between the lattice EoS and experimental data.  Similar achievements are expected soon regarding calculations of heavy quark color transport and screening.  Finally, new insights into the role of conserved charge fluctuations on the lattice and in heavy ion collisions are providing additional avenues for direct comparisons between calculations and data.

In this review, we summarize recent results in calculating the basic thermodynamic properties of high temperature QCD, with special emphasis given to calculations with physical quark masses.  These include calculations of the crossover temperature, the equation of state, and quark number susceptibilities.  We show how these results play a crucial role in understanding the properties of the quark gluon plasma created in heavy ion collisions.  We also provide a brief review of the current status of heavy quark color screening and transport, for which physical quark mass calculations are not yet in reach, and we discuss implications for future progress.  In Section~\ref{sec:actions} we describe the gluon and fermion actions used in recent QCD thermodynamics calculations and recent results are presented in Section~\ref{sec:results}.   In Section~\ref{sec:summary} we summarize recent results for color screening and transport, and in Section~\ref{sec:conclusion} we offer conclusions and discuss future prospects. 

This review is aimed primarily at students entering the field and scientists outside the field who are interested in recent results and future possibilities for lattice QCD calculations in thermodynamics.  With this goal in mind, we anticipate that experimentalists and non-lattice gauge theorists in relativistic heavy ion physics will find this review especially useful.

\section{LATTICE QCD ACTIONS AND PROPERTIES}
\label{sec:actions}

The study of QCD thermodynamics begins with the partition function,
expressed as a path integral over the classical Euclidean action
separated into fermionic and gluonic components,  
\begin{equation}
Z(T,V) = \int DA_\mu D\overline{\psi}D\psi\exp(-S^E_{QCD})
\end{equation}
\begin{equation}
S^E_{QCD} = \int_0^{1/T} dt \int_V d^3x \left[ \frac{1}{4} F^{a}_{\mu\nu}(x,t) F^{\mu\nu}_{a}(x,t) 
+ \sum_{f=1}^{n_f} \overline{\psi}^\alpha_f (x,t) \left( D^E_{\alpha\beta} - m_f \delta_{\alpha\beta}  \right) \psi^\beta_f (x,t) \right],
\end{equation}
for temperature, $T$, 
with the sum over $n_f$ different quark flavors and implicit
sums over double-indices for color degrees of freedom of
gluons ($a=1,..N_c^2-1$) and quarks ($\alpha,\beta=1,..N_c$).
In the discretized (lattice) formulation, the integrals are replaced
by sums over fermion fields occupying each 
lattice site and compactified gauge variables on the links connecting fields on
neighboring sites.  The sums are performed over $N^3_\sigma$ spatial
and $N_{\tau}$ temporal steps, related to the temperature,
$T=1/(aN_\tau)$, where $a$ is the lattice spacing.  
In many thermodynamic calculations it turns out that an aspect ratio
for $N_\sigma/N_\tau$ of 3 or 4 is already close to the infinite volume,
thermodynamic limit.  Therefore, a given lattice calculation is
referred to by specifying the number of temporal steps, $N_\tau$.  
The relative contributions of fermionic and gluonic terms vary with
the observable and temperature range.  For the trace anomaly from
which the EoS is derived, the gluonic term contributes 80\% near the
crossover, increasing to 90\% by 400~MeV.  However, the physical
contributions from quarks and gluons do not separate so cleanly.  For
example,  the fermion term vanishes for massless quarks although they
certainly contribute to the EoS. 

\begin{figure}[h]
\begin{center}
\begin{minipage}[c]{0.35\textwidth}
\includegraphics[width=0.99\textwidth]{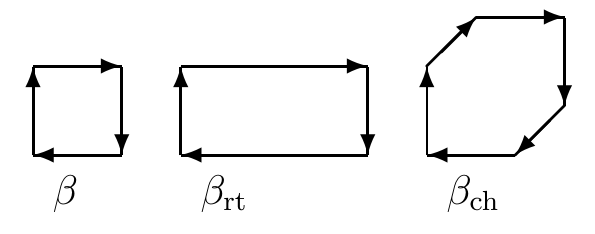}
\end{minipage}
\hspace{0.25in}
\begin{minipage}[c]{0.54\textwidth}
\includegraphics[width=0.99\textwidth]{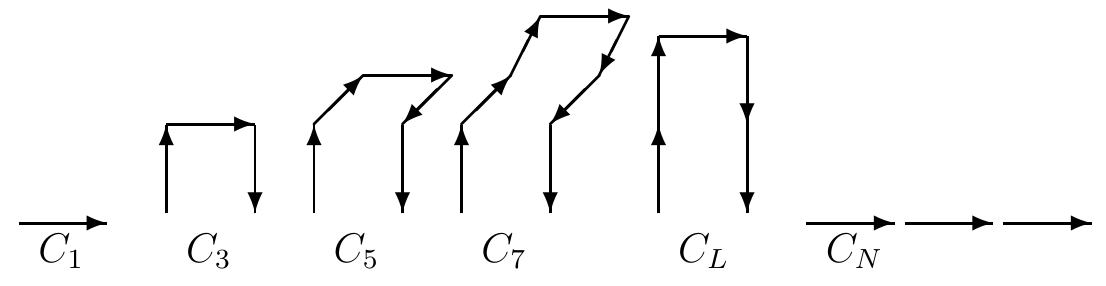}
\end{minipage}
\caption{Left: Diagrammatic representation of terms in the improved
 lattice gauge actions.  The three vector diagrams on the left depict gluonic links connecting
 neighboring space-time lattice sites.  The left most represents the
 unimproved Wilson action, referred to as a ``plaquette'', and those to
 the right represent increasing levels of improvement.  
The first diagram on the right ($C_1$) represents the unimproved staggered
fermion action and the remaining diagrams ($C_3$--$C_N$) are included
as improvements. See text for further details.}
\label{fig:imp_actions}
\end{center}
\end{figure}
Over time, additional terms have been added to the action to remove
errors of various orders in the lattice spacing, thereby improving
convergence to the continuum.  Examples of gluon terms are shown in
the left three diagrams of Figure~\ref{fig:imp_actions}. Closed paths
(Wilson loops) generate  different approximations to the continuum gauge
action term  $F_{\mu\nu}^{a}F_{a}^{\mu\nu}$.  The plaquette on the left
 ($\beta$) generates the unimproved  Wilson action.  Symanzik-improved
 gluon actions include additional terms in the action that remove
 errors of ${\cal O} (a^2)$ at tree level~\cite{Symanzik:1983kc}.
 This term is  represented by the rectangle shape
 ($\beta_{rt}$).  With the addition of the  ``chair'' ($\beta_{ch}$),
 diagram to the right, sometimes called the  ``parallelogram'', it is
 possible to eliminate, errors of ${\cal O}(\alpha_s a^2)$,
 resulting in the one-loop tadpole-improved action of L\"uscher and Weisz
 ~\cite{Luscher-Weis:1985}. Nearly all contemporary thermodynamic
 calculations employ some version of Symanzik improvement for the gluon action.  

Most of the recent algorithm improvements in lattice gauge theory have
been devoted to improvements to the fermion action.
A naive discretization of the Dirac equation for fermions introduces
additional minima in the dispersion relation at momentum component
$\pi/a$, effectively doubling the number of fermions for every
dimension, {\it i.e. $2^4$} for each flavor.  There are several
approaches to the fermion doubling problem.   The action introduced by
Wilson~\cite{Wilson:1974ji} includes an additional term that imparts a very large mass to
all but one of the fermion eigenstates, but at the expense of
explicitly breaking chiral symmetry.  Because of the important role of
chiral symmetry in finite temperature QCD, Wilson fermions are used
less often in thermodynamics,
where actions that preserve at least some aspects of chiral symmetry are preferred.

\subsection{Staggered Fermions: p4, asqtad, stout, HISQ}
\begin{figure}[h]
\begin{center}
\begin{minipage}[c]{0.59\textwidth}
\includegraphics[width=0.99\textwidth]{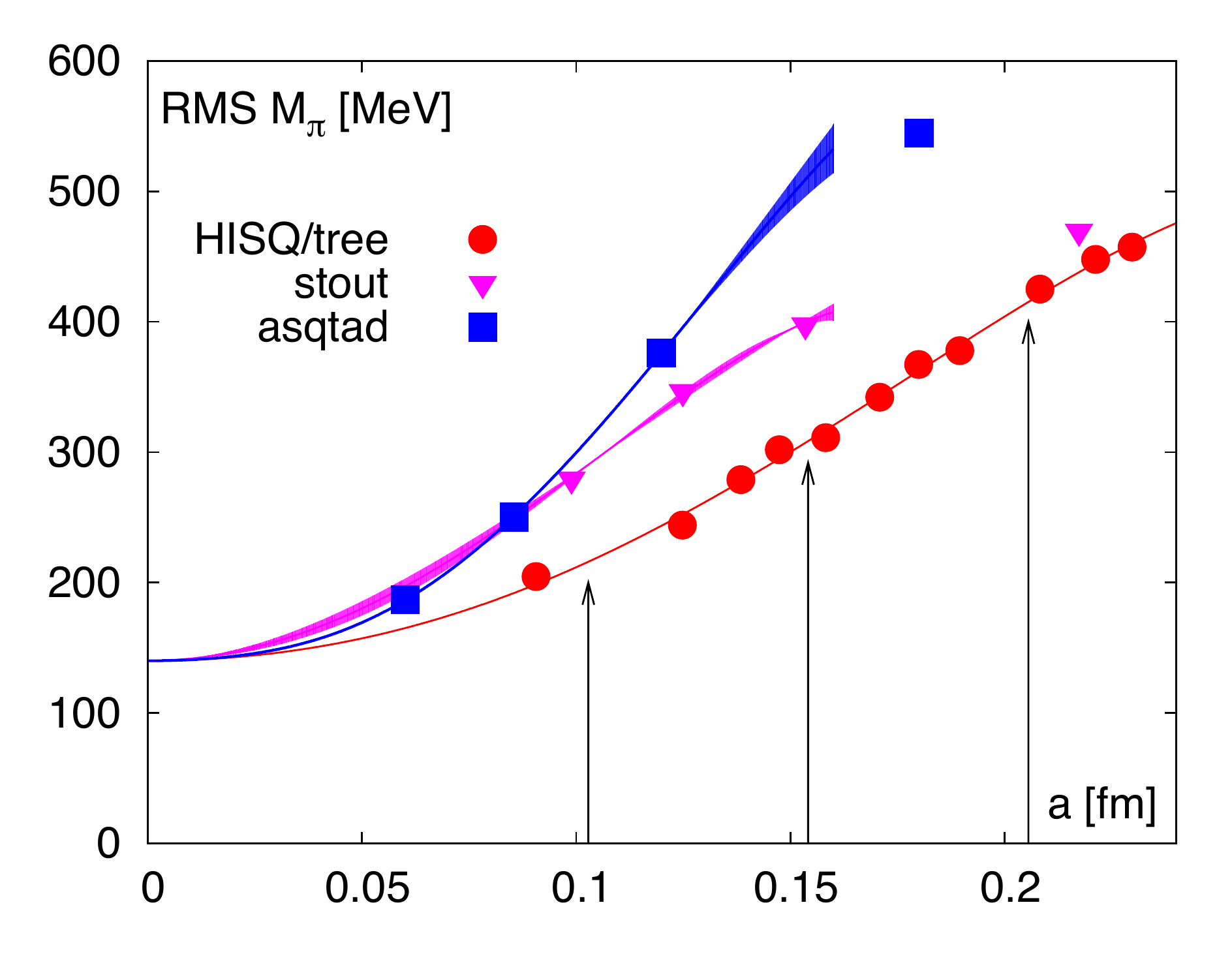}
\end{minipage}
\caption{
The root-mean-square (RMS) mass of members of the pion taste multiplet for asqtad,
stout, and HISQ/tree actions as a function of lattice spacing. These
actions employ different improvements to the staggered fermion action.
The HISQ/tree action has the lowest RMS mass over this range.}
\label{fig:pion_rms_mass}
\end{center}
\end{figure}

The staggered fermion formulation exploits an exact lattice symmetry
to reduce the number of unwanted doublers in the naive action from
sixteen to four per flavor.  In effect, the Dirac spinor
components are interleaved on alternating lattice sites.  For each flavor 
this introduces an additional, unphysical quantum number
referred to as ``taste''.   A part of the chiral symmetry on each site
is preserved, and full chiral symmetry is restored in
the continuum limit.  In this limit, the  taste degrees of freedom are
$SU(4)$-symmetric, and the correct number of degrees of 
freedom is restored by taking the fourth root of the fermion determinant
for each flavor.  The difficulty arises for finite lattice spacing, 
where the taste symmetry is broken.
The impact of taste splitting appears in the light meson spectrum.
Because a remnant of chiral symmetry is preserved, the lightest pion
behaves as the traditional Goldstone boson whose mass vanishes in the
limit of zero quark mass.  The remaining 15 members of each taste
multiplet have seven partially degenerate masses, but in the limit of zero lattice
spacing, all taste-symmetry-violating splittings vanish.  
This situation is illustrated in Figure~\ref{fig:pion_rms_mass} which shows
the root-mean-square (RMS) value of pseudo-scalar meson masses
(would be pions)
for several different staggered fermion actions to be explained below.
For physical values of the light quark masses, the different masses
from taste splitting converge towards
the pion mass in the continuum limit.

Improved fermion formulations include additional terms to reduce
lattice artifacts (``lattice cutoff effects''), one manifestation of which are the taste
splittings.  The Symanzik improvement strategy adds terms to the
fermion action so that, in the continuum limit, it still reduces to
the correct Dirac action, but errors of ${\cal O} (a^2)$ or higher are
eliminated.  This is the philosophy of the asqtad~\cite{Bernard:2007hl} 
and p4~\cite{Cheng:2008hx}
actions, which have been used extensively for QCD thermodynamics.
Diagrammatic examples of the asqtad improved
 fermion action are labeled ($C_1$--$C_N$) in
Figure~\ref{fig:imp_actions}.  The terms depict 
 fermion hopping between next-neighbor and third-neighbor sites. The
 paths of gluonic links define the gauge connection between fermion
 fields on neighboring sites.  The single $C_1$ link on the left yields the
 unimproved staggered action with discretization errors of 
 ${\cal O} (a^2)$.  The five other path shapes are needed to eliminate those
 errors completely, leaving errors of ${\cal O} (a^4)$ and ${\cal O} (\alpha_s
 a^2)$.  The p4 action is similar, but replaces the three-link term
 with a ``knights move'' term and includes other small
 changes~\cite{Cheng:2008hx}.

Further improvements are possible.  The asqtad and p4 modifications
reduce taste symmetry breaking by suppressing hard gluon exchanges
that cause transitions between tastes.  They do this by smoothing the
local gluon field experienced by the fermion.  The highly improved
staggered quark (HISQ) action has a still higher level of smoothing.
In a different approach, rather than attempting to achieve an exact
cancellation of terms of ${\cal O} (a^2)$, the Budapest-Wuppertal
group smoothed the gluon field ``stout smoothing''~\cite{Aoki:2006ge}
before coupling it to otherwise unimproved staggered fermions.

Another important measure of improvement, particularly relevant at
higher energies where quark degrees of freedom predominate, is the
extent to which the quark dispersion relation is accurately
represented.  The asqtad, p4, and HISQ actions include terms that 
eliminate ${\cal O} (a^2)$ corrections to the quark dispersion relation.  
Thus, they are expected to perform well at high energies.  The stout
action does not include these additional terms, opting instead
to perform calculations with less improvement at smaller lattice spacings.  However, in the
continuum limit all cutoff effects disappear and the various staggered actions
should all agree.

\subsection{Chiral Fermions: domain wall and overlap}
\begin{figure}[h]
\begin{center}
\includegraphics[width=0.59\textwidth]{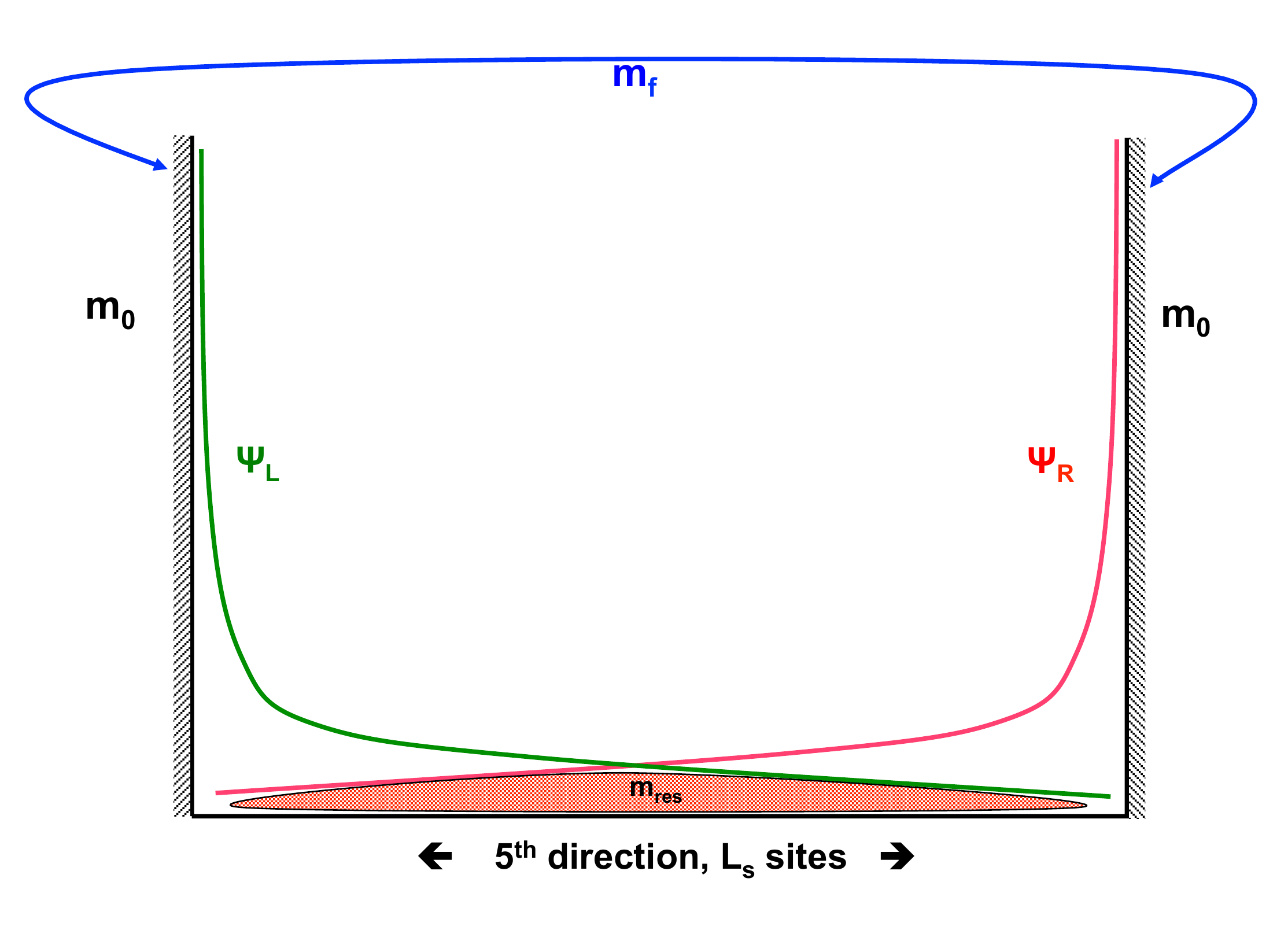}
\caption{Illustration of fifth dimension of the domain wall fermion
  action, in which the left ($\Psi_L$) and right-handed ($\Psi_R$) chiral states are
  exponentially bound to opposing walls separated by a distance of $L_s$
  lattice sites. The red region indicates the small overalp that
  occurs at finite $L_s$ and is responsible for mixing the chiral
  states and producing a residual mass $m_{\rm res}$. An explicit
  input mass $m_f$ that directly couples the two chiral states is
  introduced to provide explicit control over the mass. The ``height''
of the domain wall is denoted by $m_0$.}
\label{fig:dwf}
\end{center}
\end{figure}

To enable a fully chiral treatment of fermions on the lattice one
must be able to independently rotate the
two chiral states even at finite lattice spacing. Domain wall fermions
(DWF) achieve this by binding the two chiral states to opposing domain walls in a fictitious
fifth dimension~\cite{Kaplan:1992hb,Furman:1995we,Vranas:1998vb}.  
Figure~\ref{fig:dwf}
illustrates how the left and right handed chiral states are
exponentially localized on opposing walls, thereby achieving chiral
symmetry even at finite lattice spacing. Chiral symmetry is now
restored separately from Lorentz symmetry.  The former is fully recovered at the
$L_s \to \infty$ limit ($L_s$ is the size of the fifth dimension)
while the other at the $a \to 0 $ limit. This eliminates the taste
splitting that exists for the staggered fermion formulations and
instead gives rise to exactly 3 light pions, enabling the study of
subtle features of chiral symmetry restoration.  However, the extra
dimension brings with it an additional computational cost which
increases linearly with $L_s$.

As shown in Figure~\ref{fig:dwf} the left and right chiral wave functions
have some overlap and therefore break chiral symmetry. Because they
are expected to be exponentially localized on opposite walls, this
overlap is expected to be small. However, this overlap induces an
additive mass, called 
residual mass $m_{\rm res}$, to the input quark mass $m_f$, and as a result 
$m_{\rm eff} = m_f + m_{\rm
  res}$.  
The primary challenge in applying the DWF
action to lattice QCD is in achieving small values of $m_{\rm
 res}$ while controlling the overall computational cost that grows
with $L_s$.  This is especially true for thermodynamic calculations,
which require zero-temperature calculations on large volumes in order
to eliminate divergent vacuum contributions.
In the domain wall formulation, the gauge links are replicated on each
four-dimensional slice while the links in the fictitious fifth
dimension are all set to the unit matrix.  This allows the
construction of a transfer matrix T and corresponding Hamiltonian $H_4$.
Such a Hamiltonian framework is called the overlap
formalism~\cite{Narayanan:1993aa}.  Variants of the overlap 
formalism have been developed that make it suitable for numerical
simulations~\cite{Neuberger:1998ua,Ginsparg:1982va, Luscher_GWF} 
and have been used
in various Lattice QCD studies. These variations have  different
technical properties and problems than DWF but are exactly identical
at the infinite $L_s$ limit.  

In the DWF approach, the localization of the two chiral
components is controlled by the eigenvalue spectrum of the transfer
matrix Hamiltonian (see for example
\cite{edwards_heller_narayanan_flow}). Since the states on the 
opposite walls are related by $\exp(-L_s \times  H_4(m_0))$ their
localization depends on the eigenvalue spectrum of $H_4(m_0)$.
The lattice QCD simulation generates an ensemble of gauge field
configurations, and the eigenvalue spectrum of $H_4(m_0)$  will be
different for each configuration. As a result the localization and
$m_{\rm res}$ will vary from configuration to configuration.
For any gauge field configuration, $H_4(m_0)$ has the same number of
positive ($n_+$) and negative ($n_-$) eigenvalues for $m_0 < 0$.
However, as the height of the domain wall, $m_0$, is increased above zero some eigenvalue of
$H_4(m_0)$ may cross zero and change sign.  Then  the quantity $(n_+ - n_-)$, which is
the index of the Dirac operator,  would not be zero
just after the crossing occurs. It has been shown that the number and
direction of crossings is directly related to the number of instantons
and anti-instantons present in the gauge configuration
\cite{edwards_heller_narayanan_flow} and that $(n_+ - n_-)$ is  
equal, in a statistical sense, to the net (global) topological charge
of the gauge field configuration \cite{suri_pavlos_index}.  This is
of particular importance to faithfull lattice studies of the
$U(1)_{\rm Axial}$ symmetry breaking in QCD. 
However, it follows that a configuration that is
close to a topology change will have a near zero $H_4(m_0)$ eigenvalue 
and therefore poor localization and large $m_{\rm res}$. 

This limitation has recently been addressed by applying a Boltzmann
weight \mbox{$|H^2_4(m_0) + \epsilon^2|$} to the action.  The  $H^2_4(m_0)$
term provides a better localization and smaller
$m_{\rm res}$~\cite{vranas_gdwf},  and the
$\epsilon$ term allows one to control the topological charge crossings.
This method was first used for the study of QCD
thermodynamics in \cite{DSDR_sim_200MeV} and has been named DSDR 
(dislocation suppression determinant ratio) \cite{DSDR}. Because of the created
gap one is able to reach small enough $m_{\rm res}$ to produce three
degenerate pions at their physical value of about $140$~MeV and
faithfully simulate the QCD thermal transition
\cite{DSDR_sim_140MeV}.  The DWF algorithm continues to benefit from
algorithm improvements; see~\cite{Brower:2012ut} for a recent review.

\subsection{Temperature Scales and Cutoff Effects}

The lattice calculations are performed in dimensionless units, and converting 
the coupling parameter to physically meaningful temperature scale requires 
matching either directly or indirectly to an experimentally measured quantity.
Two common examples are the $r_0$ scale 
set by the $\Upsilon$ (2S-1S) mass splitting, and the meson decay constants, $f_\pi$, $f_K$.
The former is determined by setting the derivative of the static quark
potential $r^2\frac{dV_{\bar{q}q}(r)}{dr}=1.65$ for $r=r_0$.  A second
value $r_1$ for which this quantity is equal to unity is also used.
This scale is more suitable for calculations with smaller lattice spacings
as cut-off effects start becoming small at this distance and $r_1$ is
statistically more accurate than $r_0$. 
The decay constants
are more influenced by cutoff effects that also affect the trace anomaly  in the vicinity of the transition
temperature where this quantity is dominated by contribution from hadron resonances.  Therefore, when using the $f_K$ scale,  
thermodynamic quantities are more similar for different lattice spacings
and the continuum extrapolations are less severe.  Recently a new
Wilson flow scale, $w_0$ has been proposed, which makes use of
the $\Omega$ mass for setting the scale~\cite{Borsanyi:2012ik}.  At this time, with
lattice calculations performed much closer to the continuum limit,
the choice of scale 
parameter is less controversial than it once was.  Systematic errors associated
with the scale setting are now only 1--2\%, and many results are cross-checked 
using more than one scale setting.

The term ``cutoff effect'' is used to describe any error associated
with a particular lattice discretization scheme used in a calculation.
We have discussed the taste symmetry breaking of the staggered fermion action, but
there are additional corrections depending upon the observable being calculated.
Although all such artifacts are expected to disappear in the continuum limit, they
can lead to significant deviations from the scaling approximations used to perform
the continuum extrapolations.  Knowing when the scaling region has been reached
is perhaps the central challenge in any lattice calculation, and it is only truly evident
in hindsight, when an additional calculation at smaller lattice spacing produces no
significant change in the extrapolation.

\section{LATTICE QCD THERMODYNAMICS RESULTS WITH PHYSICAL QUARK MASSES}
\label{sec:results}
\subsection{Transition Order and Temperature}
\label{subsec:Tpc}

\begin{figure}[h]
\begin{center}
\includegraphics[width=0.49\textwidth]{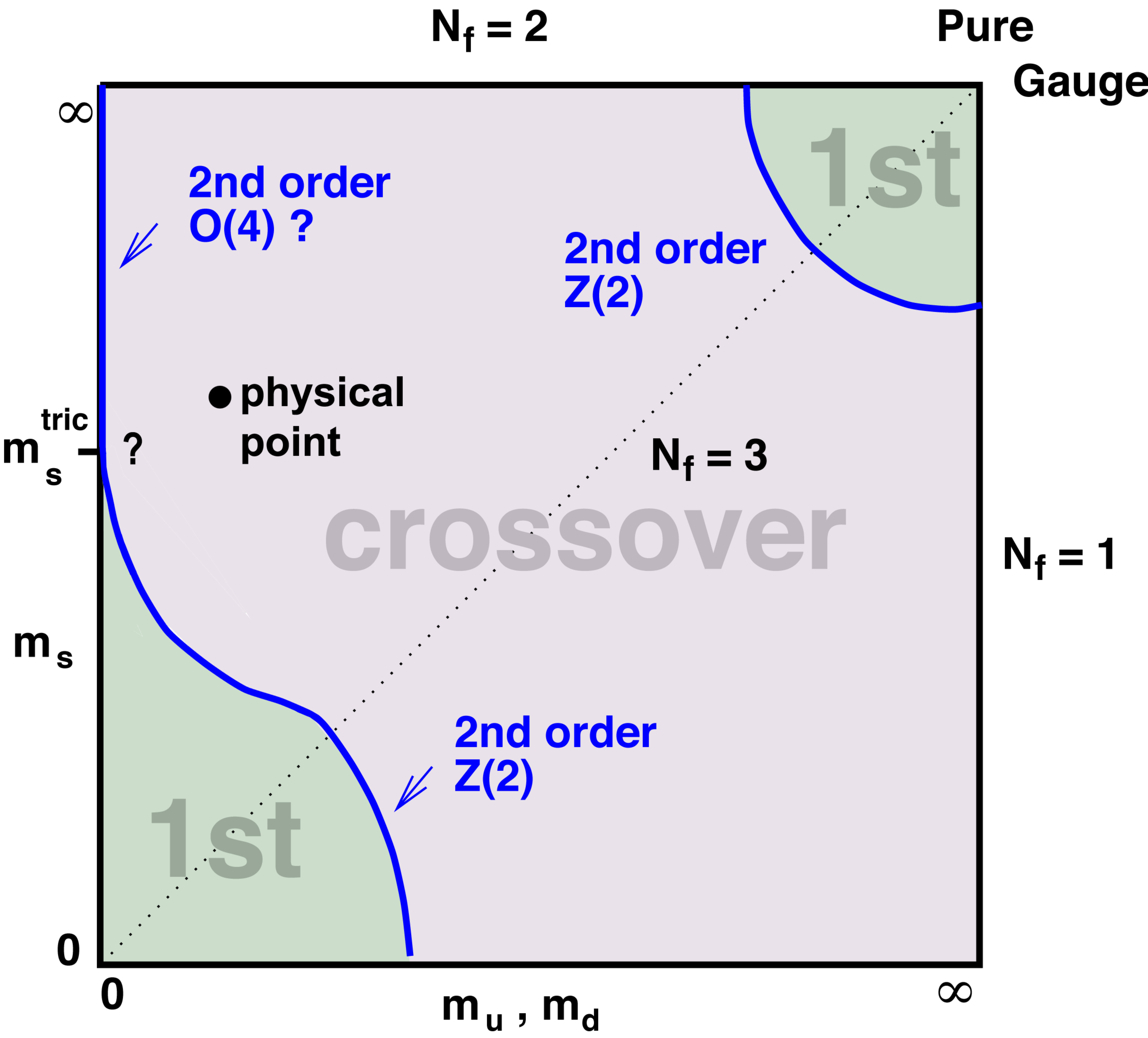}
\caption{Order of the QCD phase transition {\em vs.} strange and light quark
  mass values.}
\label{fig:sud_plane}
\end{center}
\end{figure}
The location of the QCD ``transition temperature'' has received much
attention in the heavy ion community, but a single temperature can be
defined precisely only for a true phase transition, in which the singularity in the partition function
extends to all observables.  However, understanding the nature of
the transition and defining its corresponding temperature in a
meaningful way have important implications for heavy ion collisions and
our overall understanding of QCD thermodynamics.

Figure~\ref{fig:sud_plane} illustrates the
nature of the phase transition as a function of quark mass ~\cite{Brown:1990hg,Peikert:1999cn},
in which the phase boundaries are motivated by lattice calculations and symmetry
arguments.  It illustrates the importance of having lattice calculations with physical quark 
masses, which now confirm that the transition is a crossover at the
physical point of the diagram.

\begin{figure}[h]
\begin{center}
\begin{minipage}[c]{0.42\textwidth}
\includegraphics[width=0.99\textwidth]{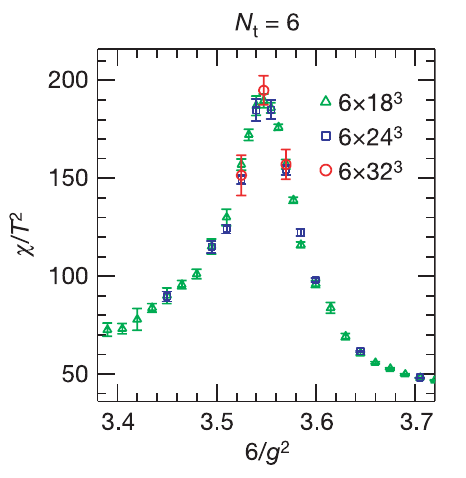}
\end{minipage}
\begin{minipage}[c]{0.56\textwidth}
\includegraphics[width=0.99\textwidth]{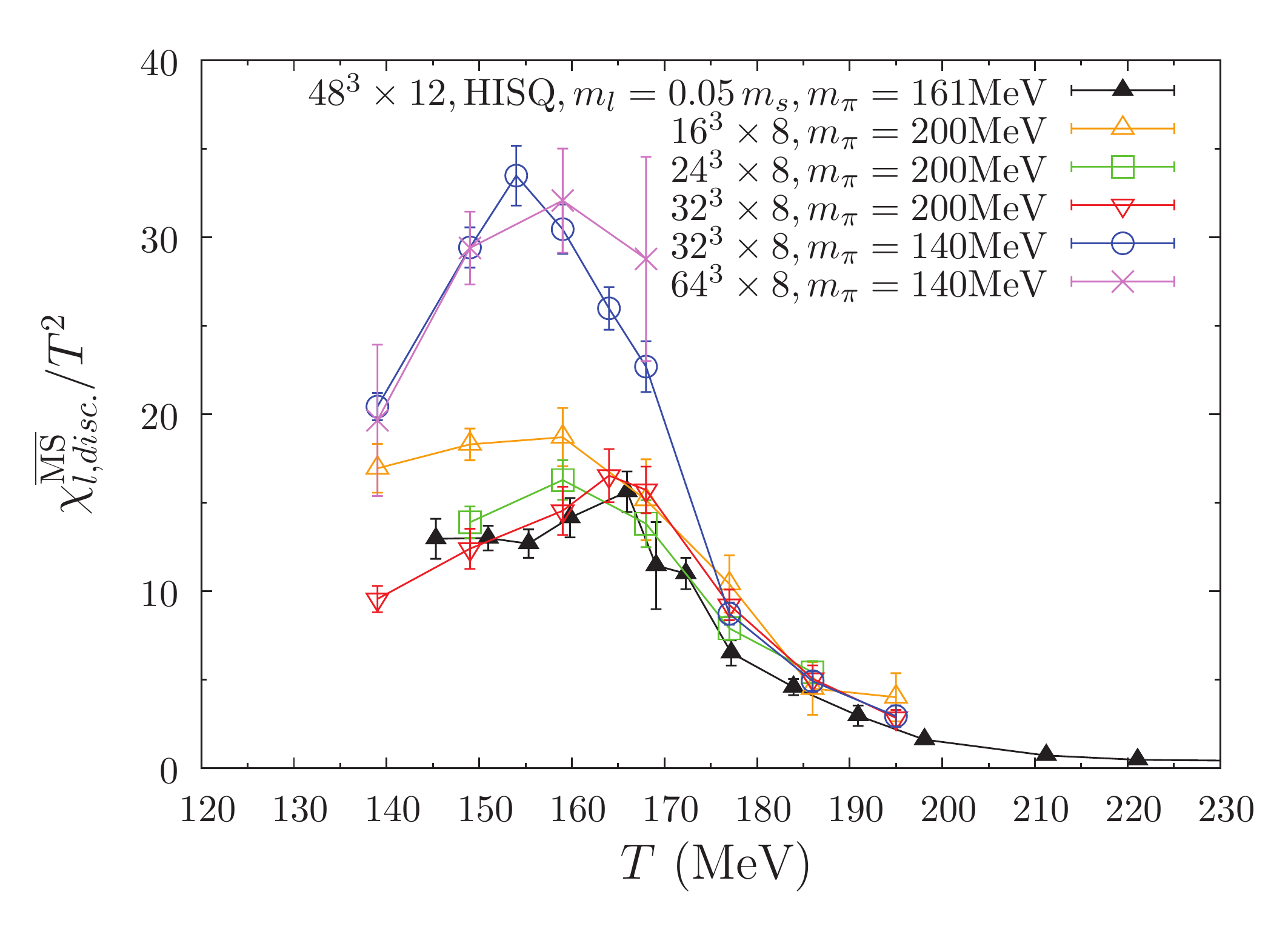}
\end{minipage}
\caption{Chiral susceptibility for several volumes for the stout
  staggered action for $N_{\tau}=6$ with physical quark mass (left),
 and the domain wall fermion action for $N_{\tau}=8$ (right) for
 $m_\pi$ values of 140 and 200~MeV.  Comparisons to the HISQ action
 with $N_{\tau}=12$ are also shown.  There is no evident change in peak height with increasing volume for
the same pion mass.}
\label{fig:chisusc_volume}
\end{center}
\end{figure}

The most direct method to probe the nature of the QCD phase transition 
is to study the derivatives of the log of the partition function with
respect to the light quark mass: the chiral condensate, $\langle
\bar{\psi}\psi \rangle$, which becomes the order
parameter in the chiral (massless) limit, and the chiral
susceptibility, $\chi_{\bar{\psi}\psi}$,
\begin{equation}
\langle \bar{\psi}\psi \rangle = \frac{T}{V} \frac{\partial
  \ln Z}{\partial m}; \hspace{1cm}
\chi_{\bar{\psi}\psi}= \frac{T}{V} \frac{\partial^2 \ln Z}{\partial m^2}.
\end{equation}
For a true phase transition the 
chiral susceptibility becomes narrower and the peak 
height increases with increasing lattice volume.
The peak height increases linearly with volume for a first order
transition and grows with critical exponents for a second order transition.  
Figure~\ref{fig:chisusc_volume} (left) shows calculations with the stout-link
improved fermion action~\cite{Aoki:2006gr}.  These calculations with
physical quark masses show no volume dependence, but calculations with
staggered fermions leave open the question of a potential systematic
error associated with the lack of a true chiral symmetry on the
lattice.  

This question has recently been answered with a calculation
using domain wall fermions, shown in the 
right panel of Figure~\ref{fig:chisusc_volume}.  Lattices with the same
lattice spacing and spatial dimensions of $32^3$ and $64^3$ have
identical values of the disconnected chiral susceptibility.

As noted, the presence of a crossover transition for physical quark
mass values complicates the definition of a transition temperature,
but many of the thermodynamic observables that develop singularities
in the chiral limit may retain some remnant of the transition in 
a steep drop or inflection point in the crossover region,
corresponding to the peak in the chiral susceptibility seen in
Figure~\ref{fig:chisusc_volume}.  These characteristics are used to
define a pseudo-critical temperature ($T_{pc}$).

\begin{figure}[h]
\begin{center}
\begin{minipage}[c]{0.49\textwidth}
\includegraphics[width=0.99\textwidth]{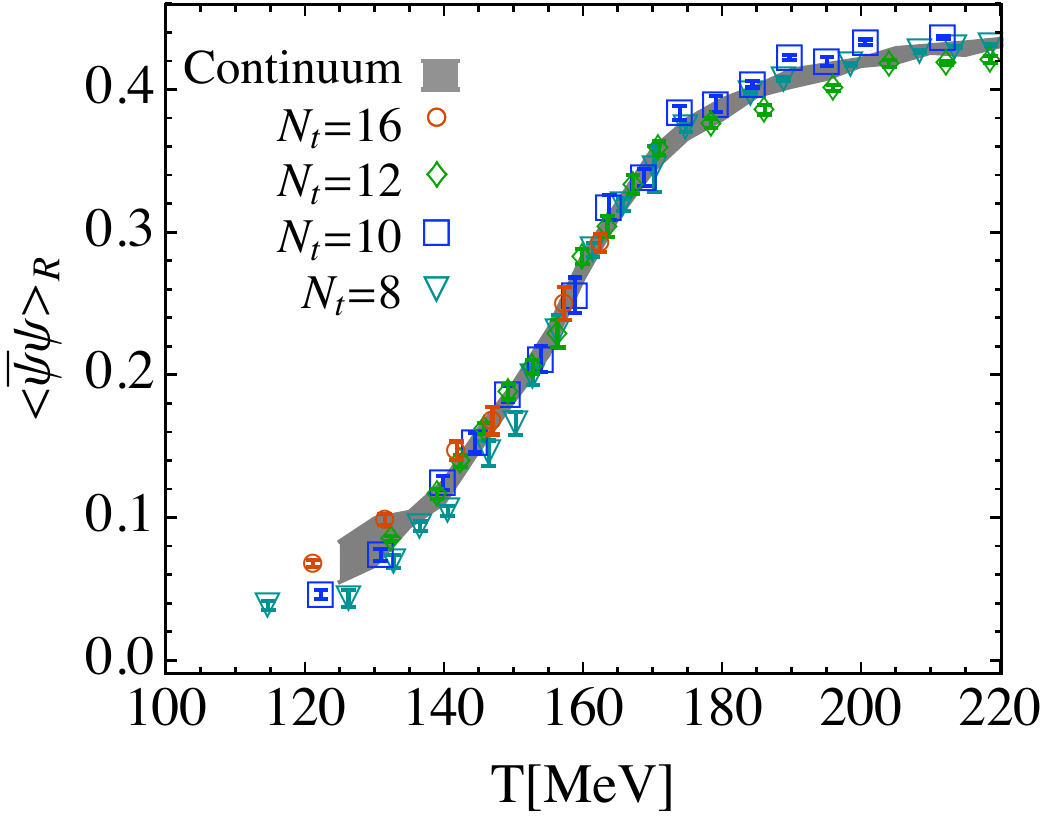}
\end{minipage}
\begin{minipage}[c]{0.49\textwidth}
\includegraphics[width=0.99\textwidth]{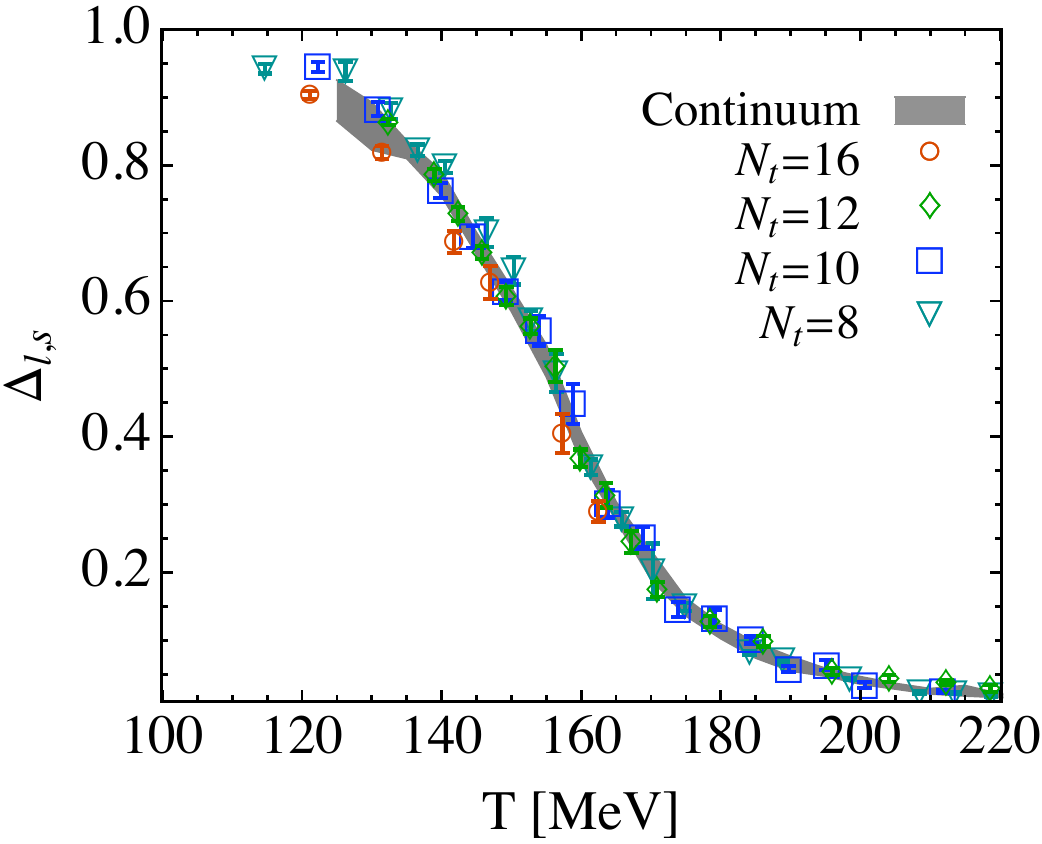}
\end{minipage}
\caption{The renormalized chiral condensate (left) and the subtracted
  renormalized chiral condensate (right) for the stout action for
  $N_\tau=8,10,12,16$ and the continuum extrapolation with physical
  quark masses.}
\label{fig:chiral_stout}
\end{center}
\end{figure}
When working with the chiral condensate, it is common to remove
lattice artifacts through subtraction and normalization.  The renormalized
chiral condensate, $ \langle \bar{\psi}\psi \rangle_R$ and the
subtracted chiral condensate, $\delta_{l,s}$, are defined as,
\begin{eqnarray}
 \langle \bar{\psi}\psi \rangle_R & = & \left[
\langle \bar{\psi}\psi \rangle_{l,T} - \langle \bar{\psi}\psi \rangle_{l,0}
\right] \frac{m_l}{T^4}, \\
\delta_{l,s} & = & \frac{
\langle \bar{\psi}\psi \rangle_{l,T} -
  \frac{m_l}{m_s} \langle \bar{\psi}\psi \rangle_{s,T} }
{\langle \bar{\psi}\psi \rangle_{l,0} -
  \frac{m_l}{m_s} \langle \bar{\psi}\psi \rangle_{s,0} },
\end{eqnarray}
where the $l$ and $s$ subscripts refer to the light and strange quark
condensates and the $T$ and $0$ subscripts refer to the finite and
zero temperature values, respectively.
Calculations with the stout action for the renormalized chiral
condensate are shown in  
Figure~\ref{fig:chiral_stout}, with the renormalized chiral condensate
shown on the left, and the subtracted chiral condensate shown on the
right~\cite{Borsanyi:2010gc}.  Results are shown for calculations with
physical quark masses for  $N_\tau=8,10,12,16$ and the continuum using
the $f_K$ scale.  Fits to the inflection point in $ \langle
\bar{\psi}\psi \rangle_R$ yield $T_{pc} = 155 \pm 2 \pm 3$~MeV, and
similar fits to $\delta_{l,s}$ yield a value of $T_{pc} =  157
\pm 3 \pm 3$~MeV.  One can also obtain $T_{pc}$ by fits to the peak
in the chiral susceptibility.  In this case the authors fit to $\chi_{
  \langle \bar{\psi}\psi \rangle}/T^4$, obtaining a value of
$T_{pc}=147 \pm 2 \pm 3$.

The HotQCD collaboration has produced a similar result, $T_{pc} =
154 \pm 8 \pm 1$~MeV, using a method that relates the
pseudo-critical behavior of the chiral susceptibility more directly
to universal properties of the critical behavior of the
true phase transition in the
chiral limit.  This is accomplished through
the use of $O(N)$ scaling relations, which enable one to parameterize the
behavior of the chiral condensate in the vicinity of the phase
transition.  The scaling relations are referred to as $O(N)$
because observations of both $O(4)$ and $O(2)$ scaling are possible.
QCD belongs to the $O(4)$ universality class in the limit
of vanishing light quark masses and a sufficiently large strange quark mass,
but the staggered fermion action retains only an $O(2)$ global symmetry.
The universal scaling relations and critical exponents for 
three dimensional, $O(N)$ symmetric models are well known and were
first exploited for a discussion of the QCD phase 
diagram by Pisarski and Wilczek~\cite{Pisarski:1983ms}.
The parametrization of the scaling functions in a form suitable
for QCD applications~\cite{Engels:2001uq} has been refined in
recent years~\cite{Engels:2011km} and has been
previously applied to the p4
action~\cite{Ejiri:2009ey}.  


Because this approach is relatively new, and establishes a
more direct link between $T_{pc}$ and $T_{c}$ in the chiral limit, we
summarize the essential features of the scaling analysis used to
perform a simultaneous fit to the asqtad and HISQ/tree continuum 
extrapolations for $m_l/m_s=1/27$.  More details can be found
in~\cite{Bazavov:2012iu} and references therein.

\begin{figure}[h]
\begin{center}
\begin{minipage}[c]{0.49\textwidth}
\includegraphics[width=0.99\textwidth]{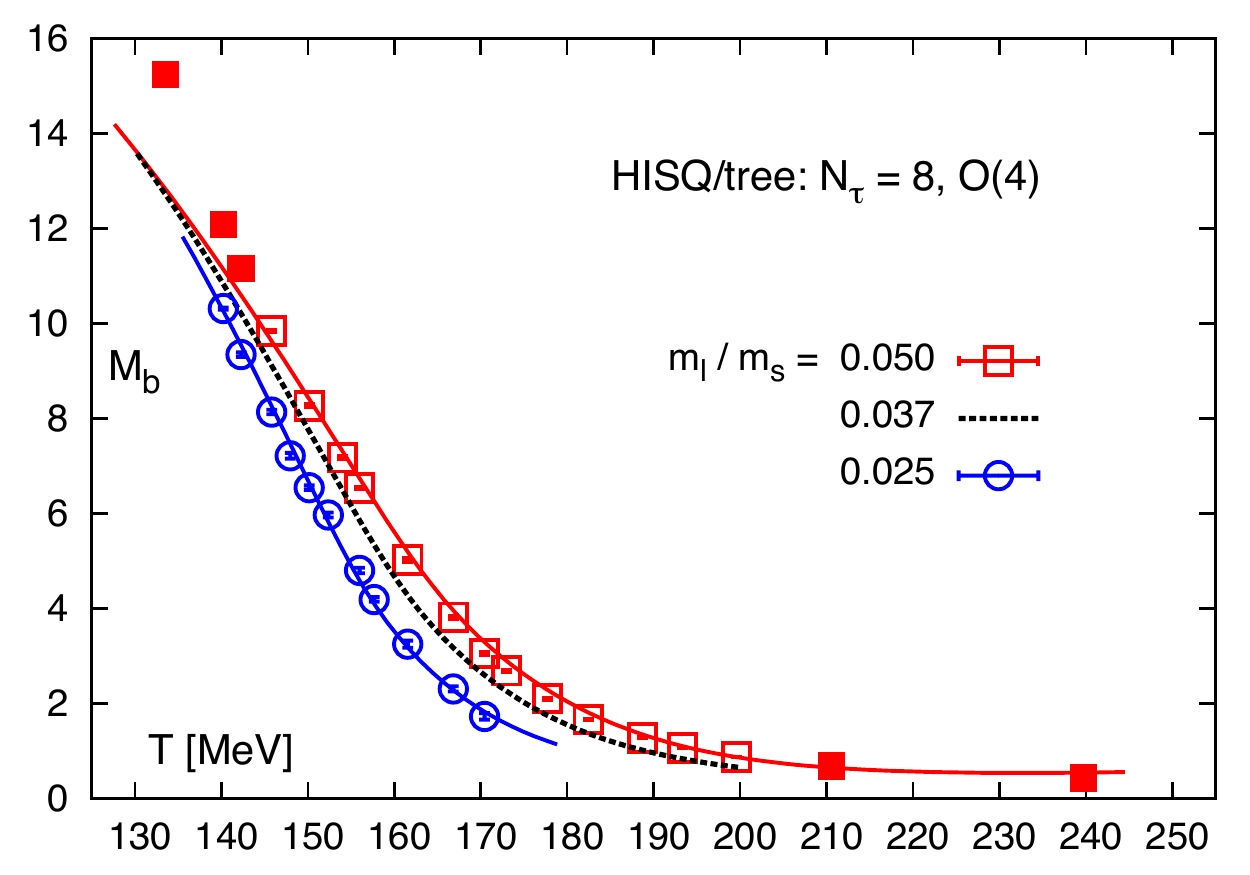}
\end{minipage}
\begin{minipage}[c]{0.49\textwidth}
\includegraphics[width=0.99\textwidth]{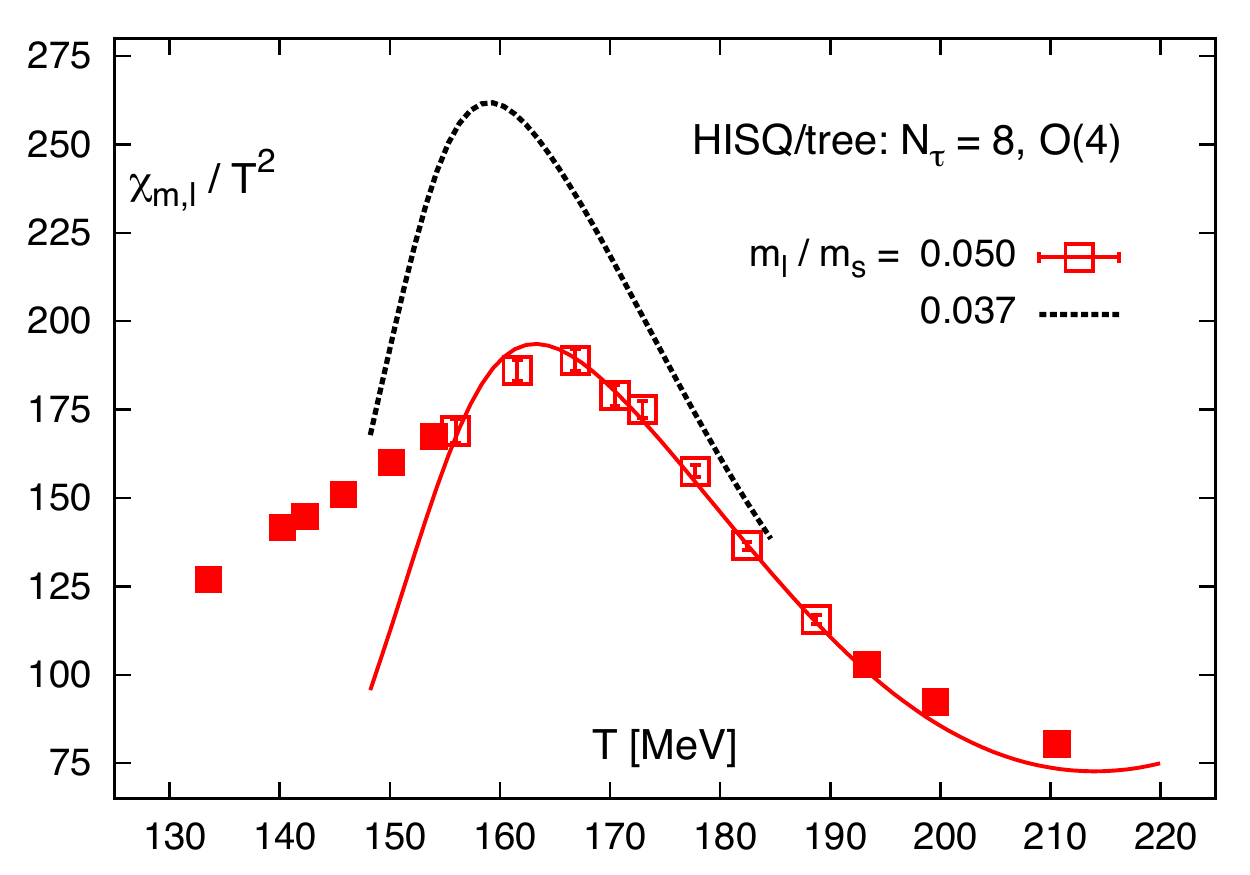}
\end{minipage}
\caption{$O(4)$ scaling fits and data for the chiral condensate $M_b$
  for the HISQ/tree action for $N_\tau=8$ (left) and the corresponding
  values of the chiral susceptibility $\chi_{m,l}$ (right).  The
  difference between the peaks of the solid red and dotted black
  curves indicate how the regular part of the transition temperature
  changes with quark mass.} 
\label{fig:chiral_hisq}
\end{center}
\end{figure}

To isolate the scaling component, one separates the singular part of the $O(N)$ phase
transition from the regular part,
\begin{equation}
f = -\frac{T}{V} \ln Z\equiv \left(\frac{m_l}{m_s}\right)^{1+1/\delta} f_{\rm singular}(z)+ f_{\rm regular}(T,m_l,m_s),
\label{free_energy}
\end{equation}
where the singular part of the free energy density is expressed in
terms of the scaling variable, $z$, which is defined in terms of dimensionless couplings related to the
temperature, $t = \frac{1}{t_0}\frac{T-T_c}{T_c}$, and the ratio of light to strange quark masses,
$h=\frac{1}{h_0}\frac{m_l}{m_s}$.  The $h$ parameter contains the quark
mass dependence and plays the role of a symmetry breaking magnetic field,
\begin{equation}
z=t/h^{1/\beta\delta}  = \frac{1}{z_0}\frac{T-T_c}{T_c} \left( \frac{m_s}{m_l} \right)^{1/\beta\delta},
\label{eq:defz}
\end{equation}
where $\beta$ and $\delta$ are the critical exponents.
The chiral condensate can then be expressed as a function of the
temperature and quark mass,
\begin{equation}
M_b \equiv
\frac{m_s \langle \bar{\psi}\psi \rangle_l^{n_f=2}}{T^4}
= h^{1/\delta} f_G(t/h^{1/\beta\delta}) + \left[ a_0 + a_1 \frac{T-T_c}{T_c} \right] h_0h,
\label{order_scaling}
\end{equation}
in which the regular part is expressed as a Taylor expansion to first order
in $t$, and the coefficients $a_0$ and $a_1$ are determined from a fit.
The parameterizations of the
scaling functions for $O(2)$ and $O(4)$ 
have been obtained from calculations with three dimensional scalar
$O(N)$ models~\cite{Engels:2011km}. 
The results from fits to the HISQ/tree
chiral condensate for $N_\tau=8$ are shown in the left panel of
Figure~\ref{fig:chiral_hisq}, and the corresponding functional forms for
the chiral susceptibility are shown on the right panel.  

\begin{figure}[h]
\begin{center}
\includegraphics[width=0.59\textwidth]{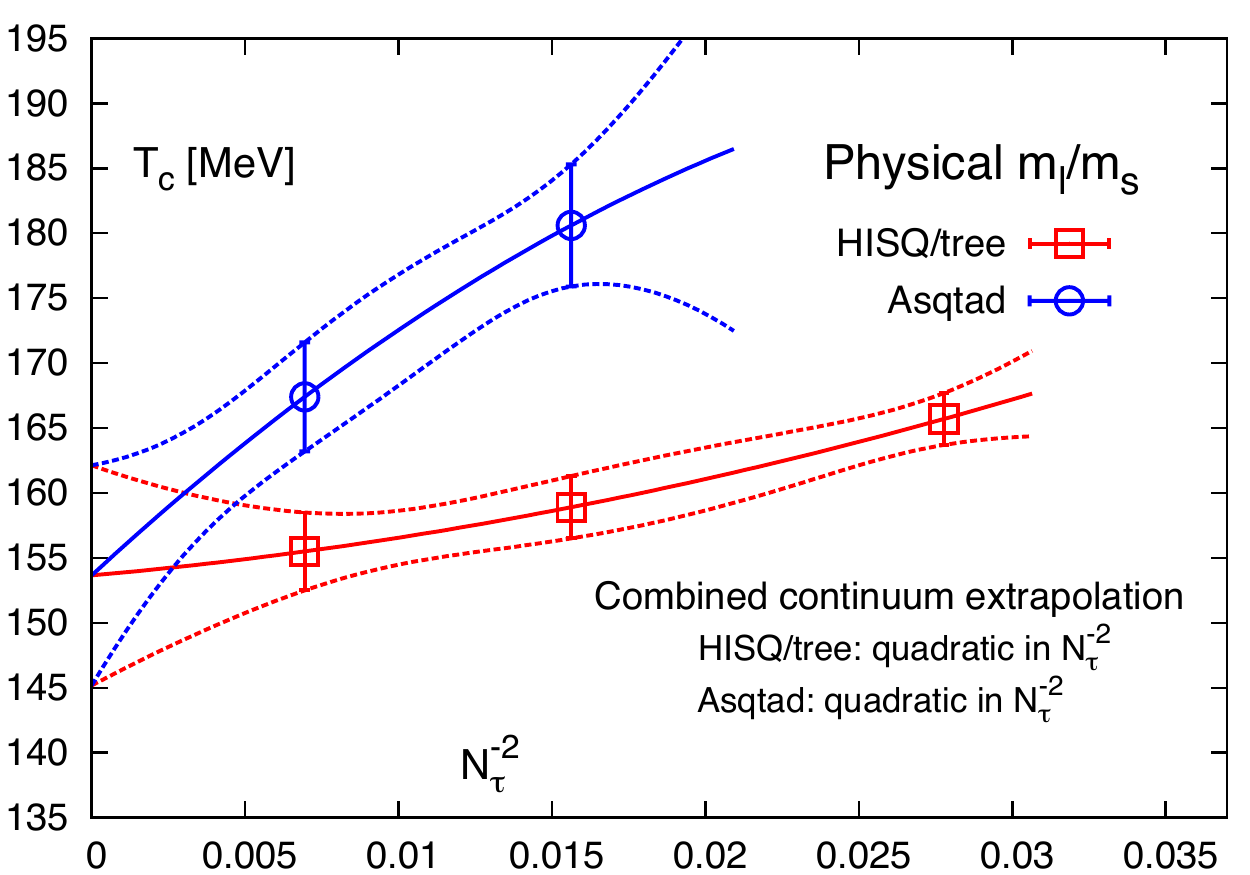}
\caption{Simultaneous quartic continuum extrapolations of the
  transition temperature for $O(4)$ scaling for both asqtad and hisq
  actions~\cite{Bazavov:2012iu}. From this analysis, the
 pseudo-critical temperature for physical light and strange quark
 masses was determined to be $154 \pm 8 \pm 1$~MeV.}
\label{fig:Tc_asqtad_hisq}
\end{center}
\end{figure}

The quark mass dependence of $T_{pc}$ is then given by,
\begin{equation}
T_{pc}\left(h\right) = T_c \left[ 1 + \frac{z_p}{z_0} \left(h\right)^{1/\beta\delta} \left(
1 -{\displaystyle\frac{a_1 t_0^\beta} {2A_p z_p {z_0}^{1-\beta} }} \,
\left( h\right) ^{1- 1/\delta  + 1/\beta\delta}  \right) \right],
\label{zero4}
\end{equation}
where $z_p$ and $A_p$ are the peak position and curvature of the scaling function
$f_\chi(z)= f_\chi(z_p)+A_p (z-z_p)^2$, which can be approximated by a quadratic
polynomial in the vicinity of the peak.
This scaling analysis was performed for two lattice
spacings for the asqtad action and three lattice spacings for
HISQ/tree.  These values were then incorporated into a simultaneous
quadratic fit in $N^2_\tau$, shown in Figure~\ref{fig:Tc_asqtad_hisq}.

\subsection{The $U(1)_{\rm Axial}$ Symmetry}
\label{subsec:U1A}

The massless QCD Lagrangian also possesses an axial,
$U_A(1)$, symmetry. This symmetry is broken due to quantum fluctuations, giving rise
to non-conservation of axial current~\cite{Adler:1969gk,Bell:1969ts} and leading to
the explicit breaking of global $U_A(1)$ symmetry through the presence of
topologically non-trivial gauge field configurations, namely the instantons
\cite{'tHooft:1976up}. With increasing temperature the density of instantons reduces
drastically due to the color Debye screening~\cite{Gross:1980br} and eventually the
$U_A(1)$ symmetry becomes exact in the $T\to\infty$ limit. Although, the $U_A(1)$
symmetry is not an exact symmetry of QCD, the magnitude of its breaking around the
chiral crossover temperature is expected to influence the nature of the chiral phase
transition. As mentioned before, for two massless flavors and in presence of
significant $U_A(1)$ breaking the chiral transition in QCD is expected to be in the
3-dimensional $O(4)$ universality class~\cite{Pisarski:1983ms,Butti:2003nu}. However,
if the $U_A(1)$ symmetry breaking becomes negligible near the chiral transition
temperature then the chiral transition in QCD can be of first order 
or of second order with the symmetry breaking
pattern $U_L(2)\times U_R(2)\to SU_V(2)$~\cite{Pelissetto:2013hqa,Grahl:2013pba}.
Therefore, determining the fate of $U_A(1)$ breaking at high
temperatures provides a more complete understanding of the chiral transition.

The issue of the axial anomaly for rooted staggered fermions is subtle and the correct
anomaly may emerge only in the continuum limit~\cite{Sharpe:2006re,Donald:2011if}.
On the other hand, the anomalous symmetry within the DWF formulation is more
straightforward~\cite{Furman:1994ky}; $U_A(1)$ is broken by the same topologically
non-trivial configurations as in the continuum and explicit lattice artifacts
appear only at order $m_{\rm res}^2$. Thus, the DWF action is a natural candidate to investigate the
temperature dependence of $U_A(1)$ breaking in QCD.

\begin{figure}[h]
\begin{center}
\begin{minipage}[c]{0.49\textwidth}
\includegraphics[width=0.99\textwidth]{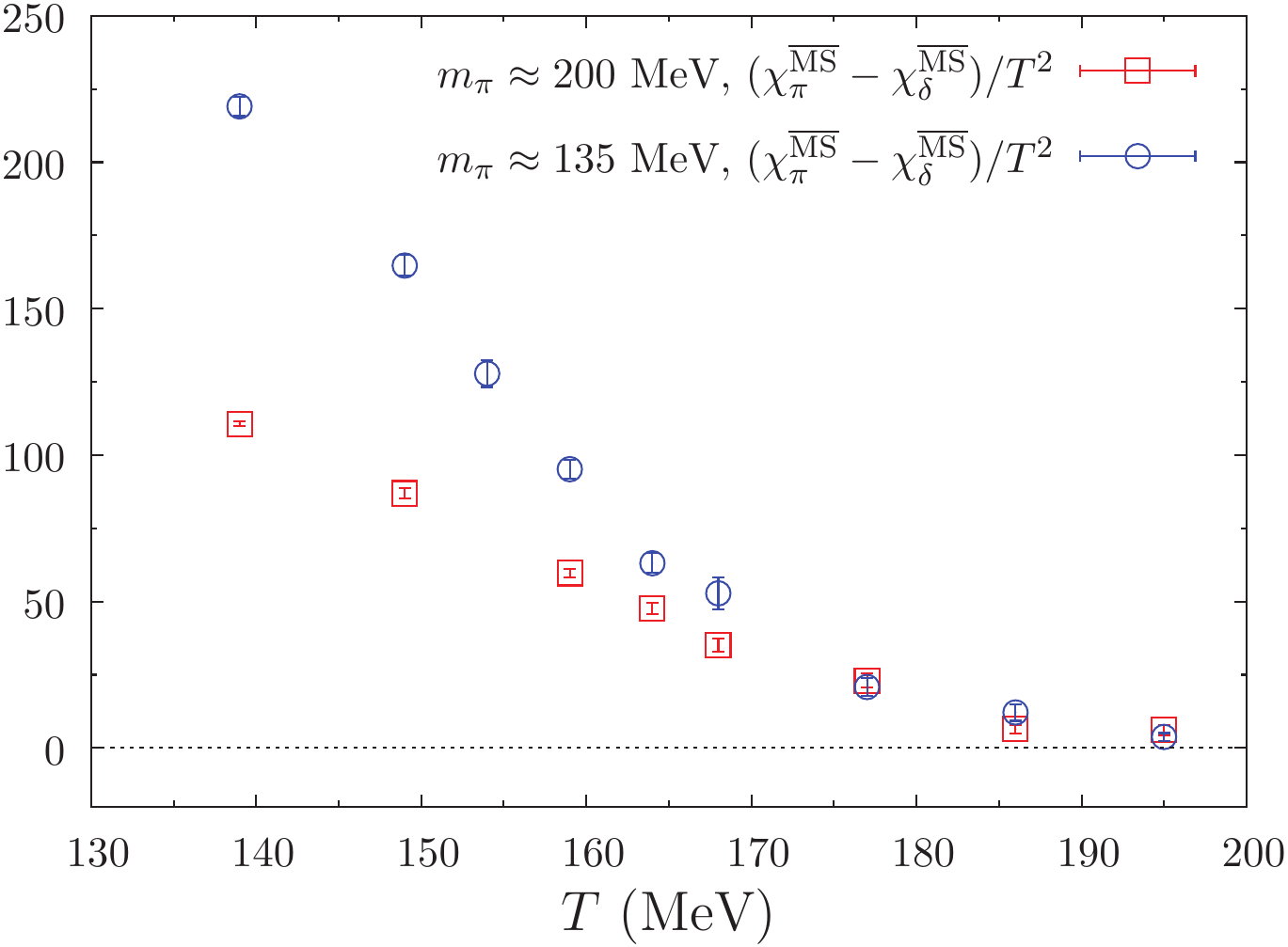}
\end{minipage}
\begin{minipage}[c]{0.49\textwidth}
\includegraphics[width=0.99\textwidth]{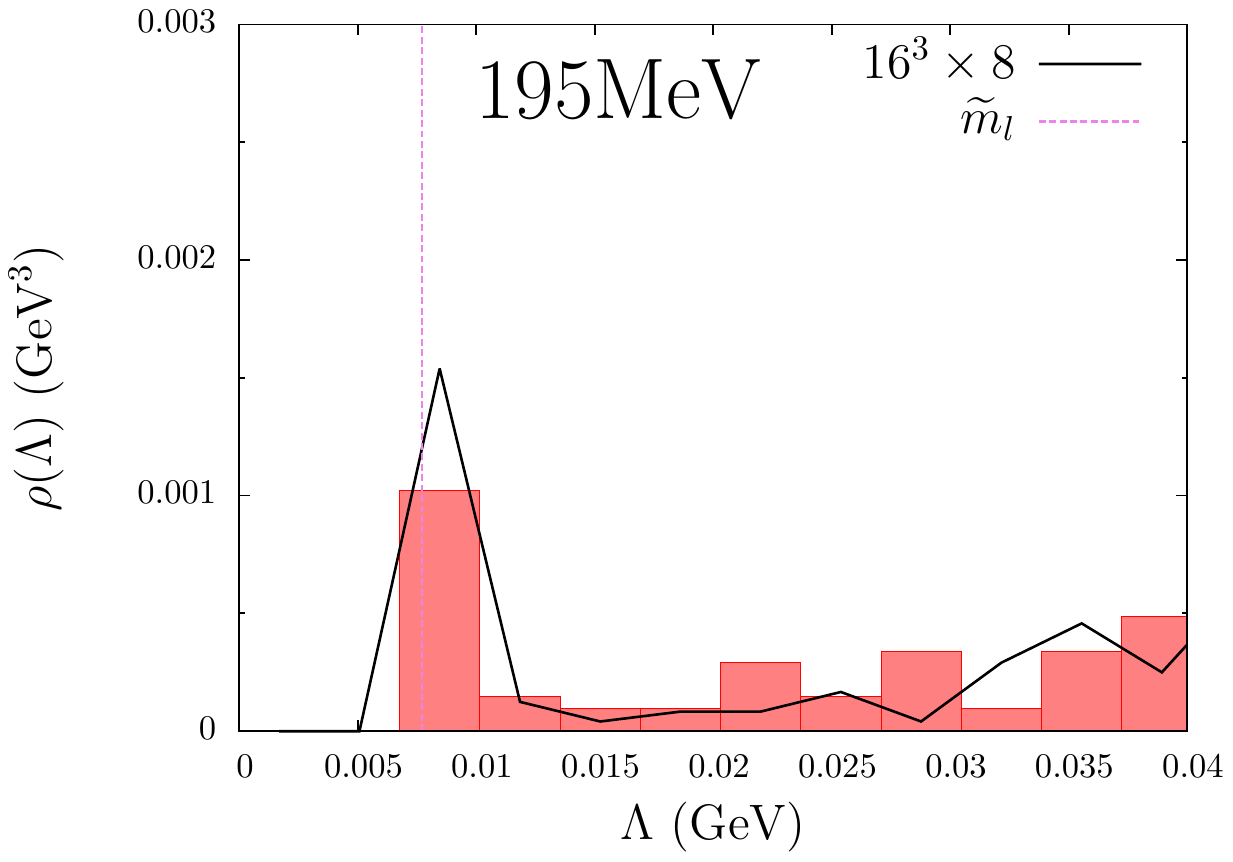}
\end{minipage}
\caption{(Left) The figure shows that the $U_A(1)$ breaking measure
$\chi_\pi-\chi_\delta$ remains non-vanishing for $T\ge T_{pc}=154(9)$ MeV
and becomes independent of quark mass for $T\gtrsim165$ MeV~\cite{Bhattacharya:2014ara}. (Right)
Infrared eigenvalue density of Dirac fermions that give rise to the $U_A(1)$ breaking
observed in $\chi_\pi-\chi_\delta$~\cite{Buchoff:2013nra}.}
\label{fig:ua1}
\end{center}
\end{figure}

For two massless flavors the pion and the isovector scalar $\delta$ $(a_1)$ mesons transform
into each other via an $U_A(1)$ rotation and the presence of an exact $U_A(1)$ will
render these mesons states degenerate. Thus, the difference of the integrated two-point
correlation functions of pion and $\delta$ meson,
\begin{equation}
\chi_\pi - \chi_\delta = \int d^4x \left[ \left\langle \pi^+(x) \pi^-(0) \right\rangle
- \left\langle \delta^+(x) \delta^-(0) \right\rangle \right ] \;,
\label{eq:chi_pi_delta}
\end{equation}
can be used as a measure of the $U_A(1)$ breaking~\cite{Shuryak:1993ee}. If the
$U_A(1)$ symmetry is exact then this quantity will vanish, and for small light quark
masses, $m_l$, the non-vanishing corrections will be of the order of $m_l^2$. This
particular measure of $U_A(1)$ breaking has been extensively studied
using the DWF action as a
function of temperature for different light quark masses and for several volumes
\cite{Bazavov:2012qja,Buchoff:2013nra,Bhattacharya:2014ara}. As shown in Figure
\ref{fig:ua1} (left), these DWF calculations clearly show that $\chi_\pi-\chi_\delta$
does not vanish around the chiral crossover temperature $T_{pc}=154(9)$ MeV and remains
independent of quark mass for $T\gtrsim165$ MeV, indicating that $U_A(1)$ may remain
broken at these temperatures even in the chiral limit.  

The $U_A(1)$ breaking is intimately related to the topology of the gauge fields and
consequently to the infrared modes of the Dirac fermions.  In the limit of infinite
volume, both the chiral order parameter, $\left\langle\bar\psi\psi\right\rangle$, and
the $U_A(1)$ breaking measure, $\chi_\pi-\chi_\delta$, can be written in terms of the
eigenvalue density, $\rho(\lambda)$, of the Dirac fermions
\begin{equation}
\left\langle\bar\psi\psi\right\rangle = \int_{-\infty}^{+\infty} d\lambda 
\frac{2m_l\rho(\lambda)}{\lambda^2 + m_l^2} \;, \qquad\mathrm{and}\qquad
\chi_\pi - \chi_\delta = \int_{-\infty}^{+\infty} d\lambda \frac{4m_l^2\rho(\lambda)}
{\left( \lambda^2 + m_l^2 \right)^2} \;.
\end{equation}
In the limit $m_l\to0$ and for $T>T_{pc}$, $\left\langle\bar\psi\psi\right\rangle$
vanishes but $\chi_\pi-\chi_\delta$ remains non-zero. This leads to an intriguing
question: what is the form of $\rho(\lambda)$ that satisfy both these requirements?
An answer to this question naturally leads to the underlying non-perturbative
mechanism of axial symmetry breaking. As shown in Figure \ref{fig:ua1}
(right), DWF
studies suggest~\cite{Bazavov:2012qja,Buchoff:2013nra} that an eigenvalue density of
the form $\rho(\lambda)\sim m_l^2\delta(\lambda)$ can largely account
for the $U_A(1)$ breaking observed in $\chi_\pi-\chi_\delta$. Such an eigenvalue density
naturally arises within a dilute instanton gas--- a gas of widely
separated, weakly interacting small instantons and anti-instantons. A more recent study using
overlap fermions, possessing even better chiral properties and an exact index
theorem, have put this underlying mechanism of $U_A(1)$ breaking at high temperatures
in a much firmer footing~\cite{Sharma:2013nva}. 

\subsection{QCD Equation of State}

\begin{figure}[h]
\begin{center}
\begin{minipage}[c]{0.49\textwidth}
\includegraphics[width=0.99\textwidth]{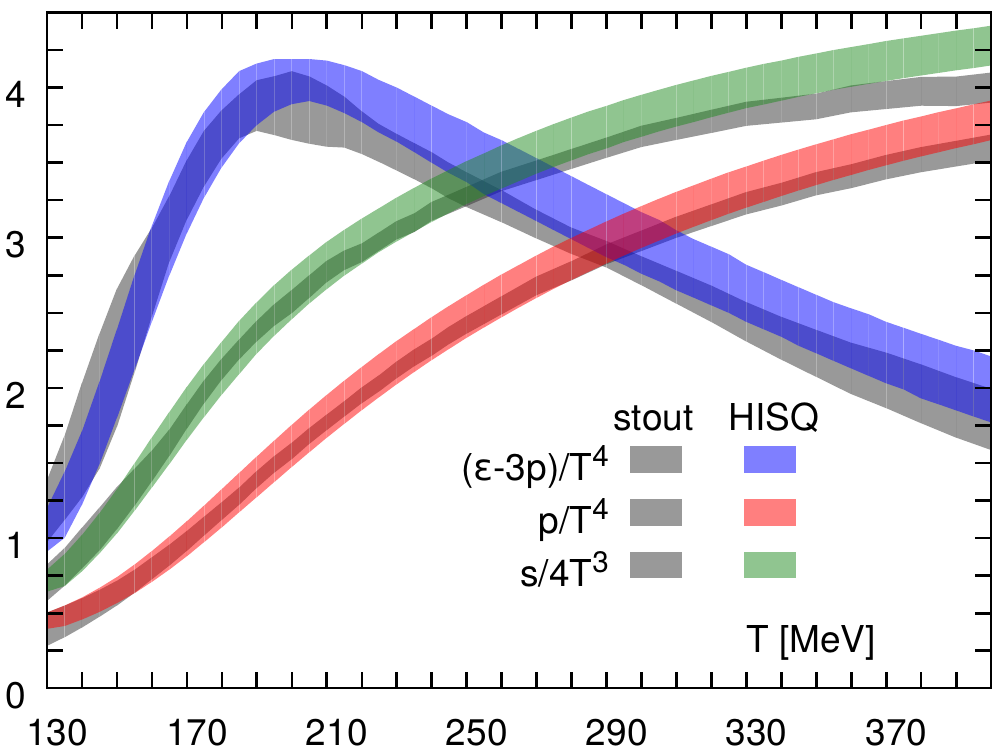}
\end{minipage}
\begin{minipage}[c]{0.49\textwidth}
\includegraphics[width=0.99\textwidth]{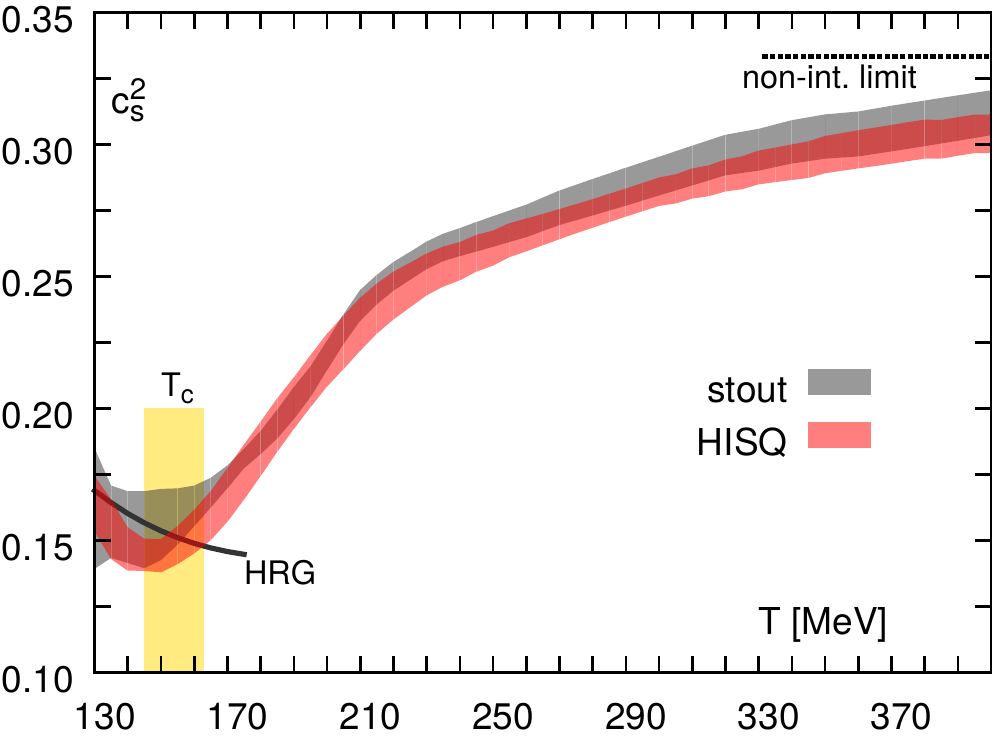}
\end{minipage}
\caption{Continuum extrapolation of the interaction measure, energy density
  and pressure for the HISQ/tree and stout actions (left), and the
  corresponding speed of sound for HISQ/tree and stout compared to the
  Hadron Resonance Gas at low temperatures.}
\label{fig:eos}
\end{center}
\end{figure}

The last few years have produced two full continuum extrapolated equation of state
calculations at or near the physical pion mass: one each from the
Wuppertal-Budapest~\cite{Borsanyi:2013up} and
HotQCD~\cite{Bazavov:2014co} collaborations.  The results, shown in
Figure~\ref{fig:eos}, are consistent within systematic errors in
the region from 130--370~MeV.  These calculations were performed with
the stout and HISQ/tree actions described in Section~\ref{sec:actions}.
The overall consistency between the two results is remarkable 
given the differences in the staggered fermion actions and analysis
methods used to extract the EoS in the continuum limit.

The equation of state is derived exclusively from the trace of the energy
momentum tensor, $\Theta^{\mu\mu}$, referred as the trace anomaly or
interaction measure because it serves to measure deviations from the
conformal equation of state, in which the energy density $\epsilon$ is
equivalent to three times the pressure $P$,
\begin{equation}
\frac{\Theta^{\mu\mu}(T)}{T^4} =\frac{\epsilon -3P}{T^4}=T\frac{d}{d T}
\left(\frac{P}{T^4}\right).
\label{theta_p}
\end{equation}
On the lattice, the trace anomaly is defined by the derivative of the
log of the partition function with respect to the lattice spacing,
$\Theta^{\mu\mu}= -\frac{T}{V}\frac{d\ln Z}{d \ln a}$, and is
evaluated separately for the gluonic and fermionic operators,
\begin{eqnarray}
&
\displaystyle
\frac{\epsilon-3P}{T^4} \equiv
\frac{\Theta^{\mu\mu}_G(T)}{T^4} +
\frac{\Theta^{\mu\mu}_F(T)}{T^4} \; , \\[2mm]
&
\displaystyle
\frac{\Theta^{\mu\mu}_G(T)}{T^4}=
R_\beta
\left[ \langle s_G \rangle_0 - \langle s_G \rangle_T \right] N_\tau^4 \; , 
\label{e3pG}  \\[2mm]
&
\displaystyle
\frac{\Theta^{\mu\mu}_F(T)}{T^4}  
= - R_\beta R_{m} [
2 m_l\left( \langle\bar{\psi}\psi \rangle_{l,0}
- \langle\bar{\psi}\psi \rangle_{l,T}\right)  \nonumber \\[2mm] 
&
\displaystyle
\qquad\qquad + m_s \left(\langle\bar{\psi}\psi \rangle_{s,0}
- \langle\bar{\psi}\psi \rangle_{s,T} \right )
 ] N_\tau^4,
\label{e3pF}
\end{eqnarray}
where $\langle s_G \rangle$ is the expectation of the gauge action,
$R_\beta=-a\frac{d\beta}{da}$ is the nonperturbative beta function,
and $R_m$ is the mass renormalization function.
The gluonic component is calculated from the difference in the expectation
values for the action at zero and finite temperatures, 
whereas the fermionic component is derived from the difference of
chiral condensates evaluated at zero and finite temperature in the light
and strange quark sectors, respectively.
Thus for the EoS calculation, each finite
temperature calculation requires an analogous computationally expensive
zero-temperature calculation on a large 
lattice (i.e. $64^4$) with equivalent coupling.  This is one reason
why EoS calculations require large, sustained computing allocations and
often take more than one year to complete.  As noted previously, 
the division between gluonic and fermionic components is not
strict, but the gluon contribution tends to dominate both the signal
and errors, especially at higher temperatures. 
Further details on the EoS the determination of $R_\beta$ and $R_m$
are given in~\cite{Cheng:2008hx,Bazavov:2014co}.

From Eq.~\ref{theta_p} it follows that the pressure is determined by
integration over the trace anomaly,
\begin{equation}
\frac{P(T)}{T^4} = \frac{P_0}{T_0^4} + \int_{T_0}^T dT'\frac{\Theta^{\mu\mu}}{T^{\prime5}},
\label{p_int}
\end{equation}
and is therefore sensitive to the value used for the pressure at the
lowest temperature.  Calculations at temperatures in the low
temperature region (significantly below $T_{pc}$) are prohibitively
expensive and in this region the hadron resonance gas is expected to
provide accurate estimates.  Furthermore, a smooth transition to the
hadron resonance gas EoS is required in order to match hydrodynamic
outputs to hadronic cascade codes when modeling heavy ion collisions.
The HotQCD collaboration uses the HRG EoS at 130~MeV as a matching
condition for the continuum extrapolation, whereas the
Wuppertal-Budapest collaboration performs an 
integration over quark mass to set the normalization of the
pressure~\cite{Borsanyi:2010hg}. 
The pressure values
calculated with each action/method differ by less than 10\% over the
full range of temperatures, well within their combined errors, and
both actions achieve a smooth transition to the HRG EoS just below the transition.

Both calculations were performed along the lines of constant physics
(LCP), in which the quark masses are held constant at physical values.  
The Wuppertal-Budapest collaboration used the ratio of the pion decay
constant to pion mass, $f_K/M_\pi$ for this, whereas the HotQCD used
the mass of the fictitious $\eta_{s\bar{s}}$ meson,
$M_{\eta_{s\bar{s}}} = \sqrt{2m^2_K-m^2_\pi}$.  The overall
temperature scale is set by Wuppertal-Budapest using $f_K$, whereas the static quark
potential and first derivative $r_{0,1}$ are used by HotQCD.  However,
both collaborations also calculate the scale with $w_0$ and obtain
consistent results that are also incorporated into their respective systematic
error estimates.

The two results also differ in the treatment of the continuum
extrapolation and the estimate of systematic errors.  The
Wuppertal-Budapest continuum 
extrapolations are performed on lattices with $N_\tau$=6, 8, and 10 with
$N_\tau$=12 included for three values of the temperature.  Above
350~MeV, only the $N_\tau$=6 and 8 were used.  The continuum
extrapolation is performed on spline fits to the data, and the
extrapolation is quadratic in the lattice spacing.

Continuum extrapolations for the HISQ/tree action were performed on
lattices with $N_\tau$=8, 10, and 12 using simultaneous quadratic fit
to splines in which the spline knot locations were included in the
overall minimization procedure.
The uncertainties were estimated by fitting 20k samples in which the
lattice calculations were allowed to vary within normal errors.  The
match to the hadron resonance gas was achieved by sampling the HRG
value at 130~MeV allowing for 10\% variation in the value and fixed
slope at that point.  Both collaborations have produced
parameterizations of their EoS calculations for insertion into
hydrodynamics models of heavy ion collisions.

Final results for the stout and HISQ/tree action trace anomaly, energy
density, and pressure are shown in the left panel of
Figure~\ref{fig:eos}.  The right panel shows the square of the speed of
sound and a comparison to the HRG calculation at low temperature.
The equation of state is an essential input for accurate modeling of
heavy ion collisions with hydrodynamic simulations.  
The equation of state is used to convert the initial Glauber or glasma
density profile to an initial temperature or entropy
profile~\cite{Schenke:2012dk}.  
Thereafter only the speed of sound enters into the hydrodynamic calculation.  
Before lattice calculations were able to provide smooth
parameterizations to the modeling community, an over-reliance on
simple formulas, such as the bag model equation of state with first
order phase transition added to the challenge of correctly modeling
the space-time distributions at freeze-out, as measured by femtoscopic
correlations~\cite{Lisa:2005cg,Pratt:2009bka}.  It was only through
the simultaneous adoption of a lattice-inspired EoS and second order
viscous terms that provided the modeling community with the tools
needed to successfully model the evolution of a heavy ion
collision~\cite{Luzum:2008hz}.

The question at this time is whether current uncertainties
are sufficient for current and future modeling needs, or whether
additional refinements are needed (at significant computational
expense).  The answer to this question is not yet rigorously known, and
depends upon current multiparameter sensitivity
studies that are just beginning~\cite{Soltz:2012rk,Novak:2013tf,Pratt:2015vb} .  The prevailing
consensus is that current uncertainties in the lattice EoS are
sufficient and that further refinements will not greatly elucidate
the dominant uncertainty in the understanding and parameterization of
the initial conditions.  A more rigorous determination of the
uncertainties needed in the lattice EoS is expected within the
next few years.

\subsection{Fluctuations, Freeze-out, and Finite Baryon Density}

As noted, lattice QCD studies have firmly established that at high
temperatures and zero baryon chemical potential normal hadronic matter
turns into a QGP
through a smooth, but rapid crossover. However, based on various 
theoretical studies it is generally believed that at large baryon densities and small
temperatures the transition from hadronic to QGP matter takes place via a
first order phase transition. The conjectured point in the temperature--baryon
chemical potential phase diagram of QCD, at which this first order transition line
meets the crossover region is known as the QCD critical (end) point
\cite{Halasz:1998qr,Berges:1998rc}. The QCD critical point is a unique point in the
QCD phase diagram beyond which the hadronic and the QGP phase coexist along the first
order line. At the QCD critical point, a second order phase transition takes place
between the hadronic and the QGP phase, resulting in long-range correlations at all
length scales. 

The large correlation length ($\xi$) associated with a nearby critical
point manifests itself through increased fluctuations. For example,
while the second cumulant of conserved 
charge fluctuations scales as $\xi^2$, the higher order cubic and quartic cumulants
grow as $\xi^{4.5}$ and $\xi^7$ respectively~\cite{Stephanov:2008qz}. Furthermore, it
has been shown that even qualitative features of higher cumulants can signal presence
of criticality~\cite{Stephanov:2011pb,Friman:2011pf,Asakawa:2009aj}. These cumulants
can also be accessed in heavy ion collisions via event-by-event fluctuations
\cite{Jeon:2003gk}. In this vein, the search for the QCD critical point in the
Relativistic Heavy Ion Collider's (RHIC) Beam Energy Scan (BES) program is
concentrated on measurements of higher order cumulants of the fluctuations
\cite{Aggarwal:2010wy,Adamczyk:2013dal,Adamczyk:2014fia}. These higher cumulants of
conserved charge fluctuations are also accessible in lattice QCD calculations. Among
several conserved charges the net electric charge is of special interest as its
fluctuations can provide a direction comparison between experimental
measurements and lattice QCD~\cite{Bazavov:2012jq}.

Although a direct lattice QCD computation at non-zero baryon ($\mu_B$), charge
($\mu_Q$) or strangeness ($\mu_S$) chemical potentials remains difficult due to the
infamous sign problem, higher cumulants of fluctuations of these conserved charges
can be computed on the lattice using the well established method of Taylor expansion
\cite{tayexp,Gavai:2003mf}. In this method one expands the QCD logarithm
of the partition function, $\ln Z$, or the pressure, $P=-T\ln(Z)/V$,
in a power series of the chemical potentials around vanishing values of the chemical
potentials. For the electric charge chemical potential
\begin{equation}
\frac {P \left( T, \mu_Q \right)} {T^4} = \sum_{n=0}^\infty
\frac{1}{n!} \: \chi^Q_n(T) \left( \frac{\mu_Q}{T} \right)^n
\;, \; \mathrm{where} \qquad
\chi^Q_n(T) =  \frac{1}{VT^3} \: \left. 
\frac {\partial^n \ln \mathcal{Z}} {\partial \left( \mu_Q / T \right)^n}
\right\vert_{\mu_Q=0}
\;.
\end{equation}
Here, $V$ and $T$ denote the volume and the temperature respectively. The
coefficients $\chi^Q_n$ are known as the generalized susceptibilities associated with
the specific conserved charge. Since these generalized susceptibilities are defined
at vanishing chemical potentials, lattice QCD simulations can be used to compute
them. In order to obtain these susceptibilities at $\mu_B\ne0$, one can
further Taylor expand in a power series of $\mu_B$ around $\mu_B=0$
\begin{equation}
\chi^Q_n \left( T, \mu_B \right) = \sum_{k=0}^\infty \frac{1}{k!} \: 
\chi^{BQ}_{kn}(T) \left( \frac{\mu_B}{T} \right)^k
\;, \; \mathrm{where} \qquad
\chi^{BQ}_{kn}(T) = \left. 
\frac {\partial^k \chi^Q_n } {\partial \left( \mu_B / T \right)^k}
\right\vert_{\mu_B=0}
\;.
\end{equation}
These generalized susceptibilities are, in turn, related to the fluctuations of the
conserved charges
\begin{eqnarray}
\chi^Q_1 \left( T, \mu_B \right) = \frac{1}{VT^3} \left\langle N_Q \right\rangle
\;, &\qquad&
\chi^Q_2 \left( T, \mu_B \right) = \frac{1}{VT^3} 
\left\langle \left( \delta N_Q \right)^2 \right\rangle
\;, \nonumber \\
\chi^Q_3 \left( T, \mu_B \right) = \frac{1}{VT^3} 
\left\langle \left( \delta N_Q \right)^3 \right\rangle
\;, &\qquad&
\chi^Q_4 \left( T, \mu_B \right) = \frac{1}{VT^3} \left[
\left\langle \left( \delta N_Q \right)^4 \right\rangle - 3 \left\langle \left( \delta
N_Q \right)^2 \right\rangle^2 \right]
\;,
\end{eqnarray}
where $N_Q$ is the net (positive minus negative) charge and $\delta N_Q = N_Q -
\left\langle N_Q \right\rangle$.  

On the other hand, heavy ion experiments measure various cumulants, such as the mean ($M$),
variance ($\sigma$), skewness ($S$), kurtosis ($\kappa$) {\it etc.}, of the
event-by-event distribution of the net charge~\cite{Aggarwal:2010wy} at a given
collision energy ($\sqrt{s}$).  These cumulants are related to the higher order
non-Gaussian fluctuations of conserved charge; as an example, for the net charge
\cite{Adamczyk:2014fia}
\begin{eqnarray}
M_Q \left( \sqrt{s} \right) = \left\langle N_Q \right\rangle
\;, &\qquad\qquad&  
\sigma_Q^2 \left( \sqrt{s} \right) = \left\langle \left( \delta N_Q \right)^2 \right\rangle
\;, \nonumber \\ 
S_Q \left( \sqrt{s} \right) = \frac {\left\langle \left( \delta N_Q \right)^3 \right\rangle} {\sigma_Q^3}
\;, &\qquad\qquad& 
\kappa_Q \left( \sqrt{s} \right) = 
\frac {\left\langle \left( \delta N_Q \right)^4 \right\rangle} {\sigma_Q^4} - 3
\;.
\end{eqnarray}

\begin{figure}[h]
\begin{center}
\begin{minipage}[c]{0.42\textwidth}
\includegraphics[width=0.99\textwidth]{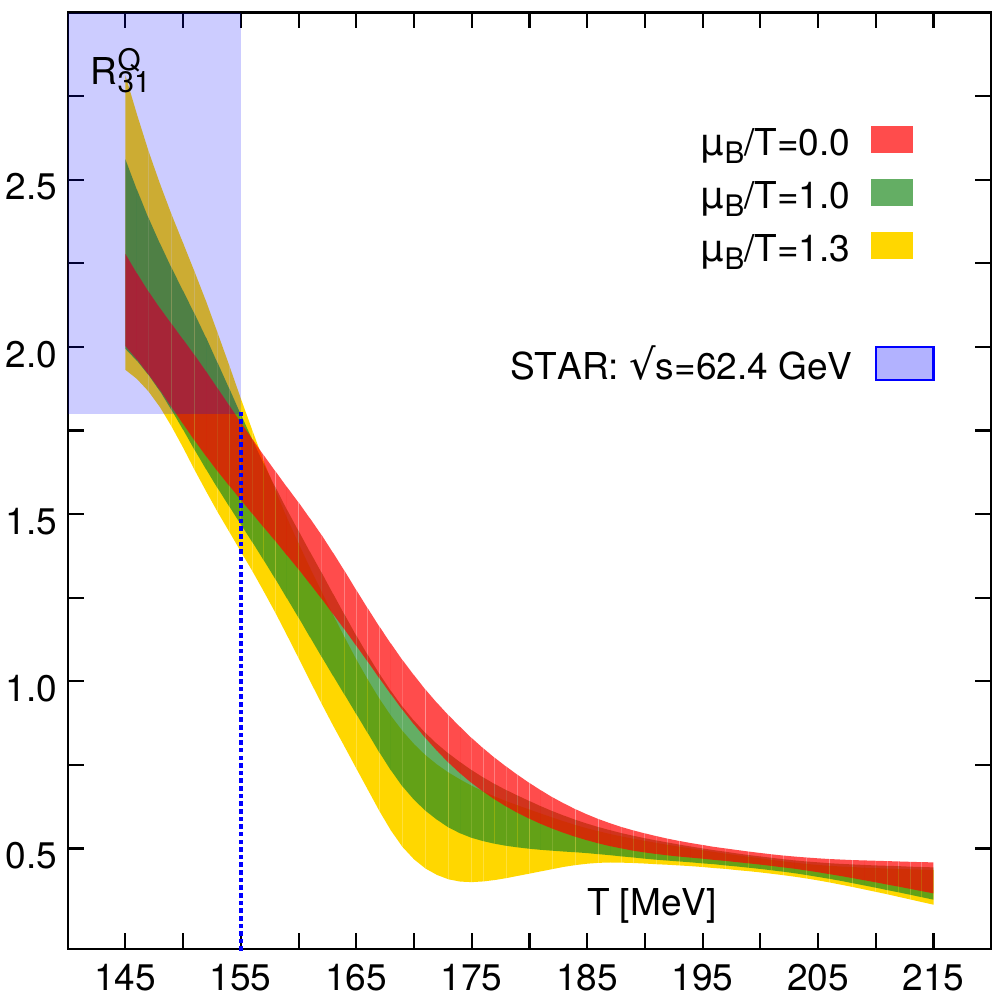}
\end{minipage}
\begin{minipage}[c]{0.56\textwidth}
\includegraphics[width=0.99\textwidth]{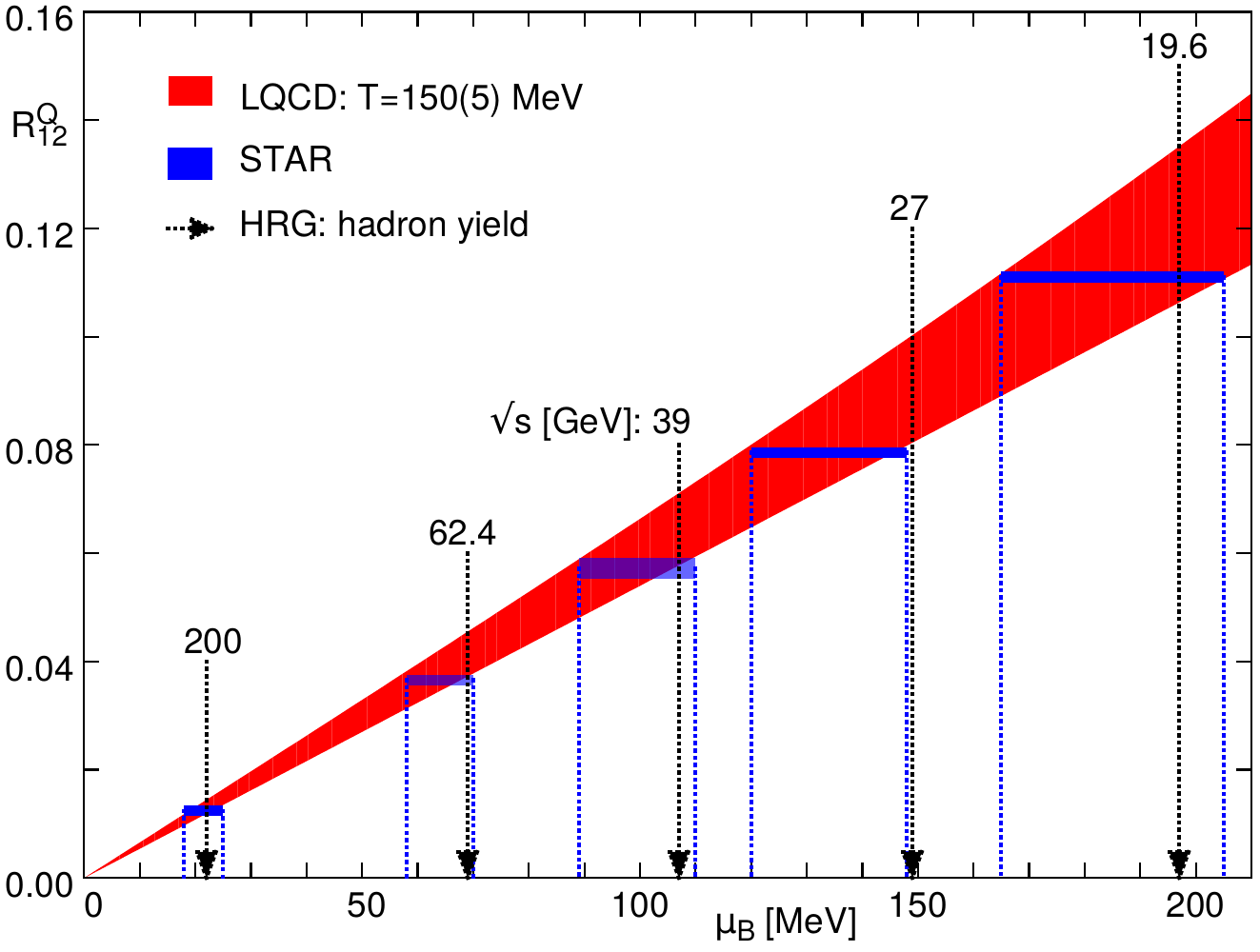}
\end{minipage}
\caption{(Left) The figure shows a comparison between the lattice QCD results
\cite{Bazavov:2012vg} for $R_{31}^Q$ and the STAR data~\cite{Adamczyk:2014fia} for
$(S_Q \sigma_Q^3)/M_Q$ at $\sqrt{s}=62.4$ GeV.  The overlap of the experimental
results with the lattice QCD calculations provides an upper bound on
the freeze-out
temperature $T^f\le155$ MeV. (Right) Lattice QCD results~\cite{Bazavov:2012vg} for
$R_{12}^Q$ as a function of $\mu_B$ compared with the STAR data
\cite{Adamczyk:2014fia} for $M_Q/\sigma_Q^2$ in the temperature range $T^f=150(5)$
MeV. The overlap regions of the experimentally measured results with the lattice QCD
calculations provide estimates for the freeze-out chemical potential $\mu_B^f$ for a
given $\sqrt{s}$. The arrows indicate the corresponding values of $\mu_B^f$ obtained
from the statistical model fits to the experimentally measured hadron yields
\cite{Andronic:2011yq}.}
\label{fig:fo_lqcd}
\end{center}
\end{figure}

Recent experimental advances in measurements of cumulants of charge fluctuations have
placed us in a unique situation where, for the first time, lattice QCD
computations at non-zero temperatures and densities can be directly confronted with
the results from heavy ion experiments through the use of appropriate volume-independent
ratios of cumulants of net charge fluctuations~\cite{Bazavov:2012vg}
\begin{subequations}
\begin{eqnarray}
\frac {M_Q \left( \sqrt{s} \right)} {\sigma_Q^2 \left( \sqrt{s} \right)} &=& 
\frac {\chi^Q_1 \left( T, \mu_B \right)} {\chi^Q_2 \left( T, \mu_B \right)} 
\equiv R_{12}^Q \;, \label{eq:R12Q} \\
\frac {S_Q \left( \sqrt{s} \right) \sigma_Q^3 \left( \sqrt{s} \right)}
{M_Q \left( \sqrt{s} \right)} &=& 
\frac {\chi^Q_3 \left( T, \mu_B \right)} {\chi^Q_1 \left( T, \mu_B \right)} 
\equiv R_{31}^Q \;. \label{eq:R31Q}
\end{eqnarray}
\end{subequations}
To obtain any information regarding the location of the critical point in the
$T-\mu_B$ phase diagram of QCD  from experimentally measured cumulants of charge
fluctuations, it is essential to relate the experimentally
tunable parameter $\sqrt{s}$ to thermodynamic parameters, namely the freeze-out
temperature $T^f$ and freeze-out chemical potential $\mu_B^f$. Recently, it has been
shown~\cite{Bazavov:2012vg} that it is possible to extract these thermal freeze-out
parameters $T^f$ and $\mu_B^f$ by comparing first principles lattice
calculations for $R_{31}^Q$ [Eq. \ref{eq:R31Q}] and $R_{12}^Q$
[Eq. \ref{eq:R12Q}], directly to their corresponding cumulant 
ratios in heavy ion collisions.  The feasibility of such a procedure has been demonstrated in
\cite{Mukherjee:2013lsa,Borsanyi:2013hza,Borsanyi:2014ewa}. A recent example of such
a comparison and subsequent determination of the freeze-out parameters are shown in
Figure \ref{fig:fo_lqcd} (see the figure caption for details)~\cite{MukherjeeCPOD14}.

\begin{figure}[h]
\begin{center}
\begin{minipage}[c]{0.49\textwidth}
\includegraphics[width=0.99\textwidth]{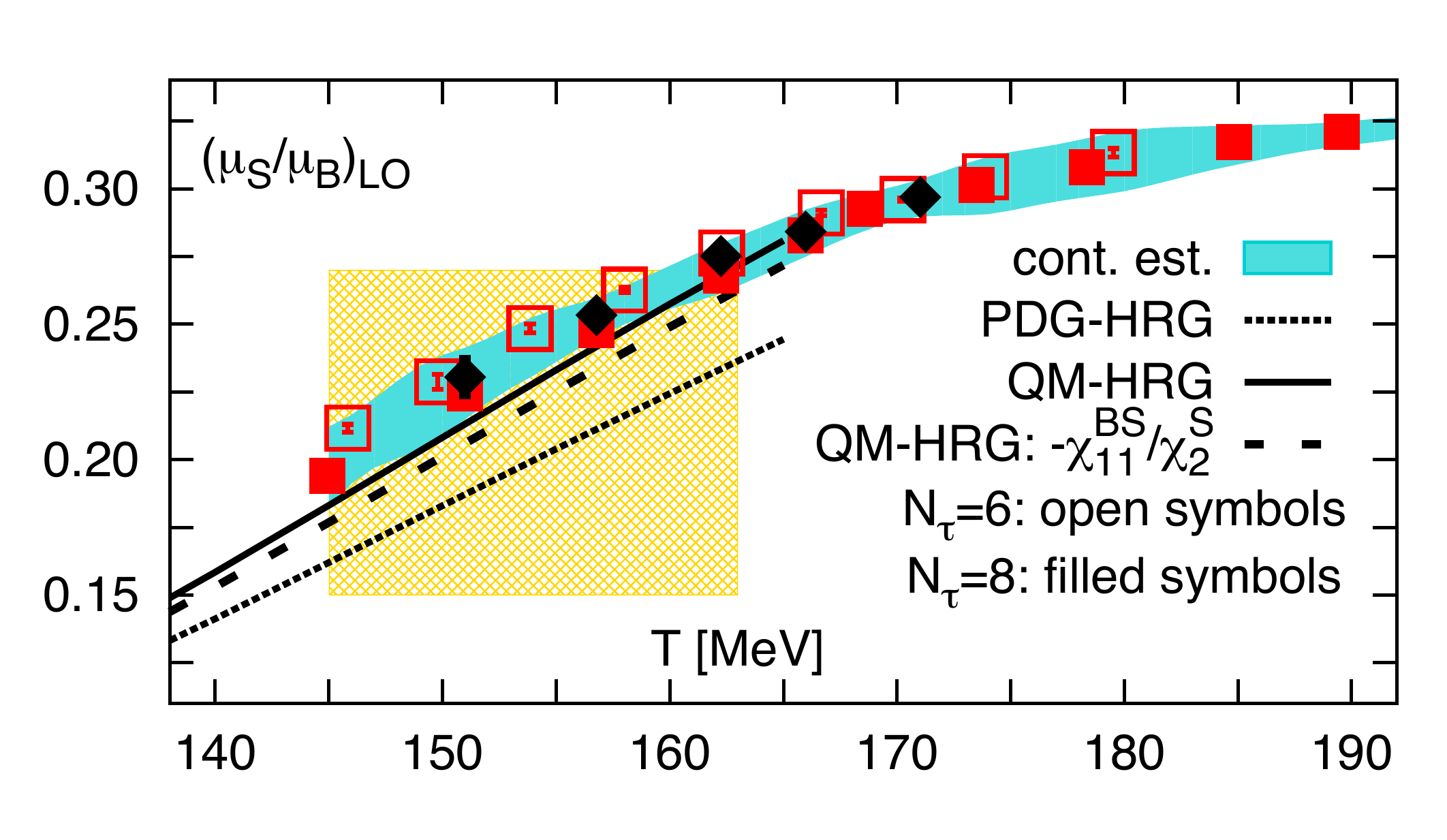}
\end{minipage}
\begin{minipage}[c]{0.49\textwidth}
\includegraphics[width=0.99\textwidth]{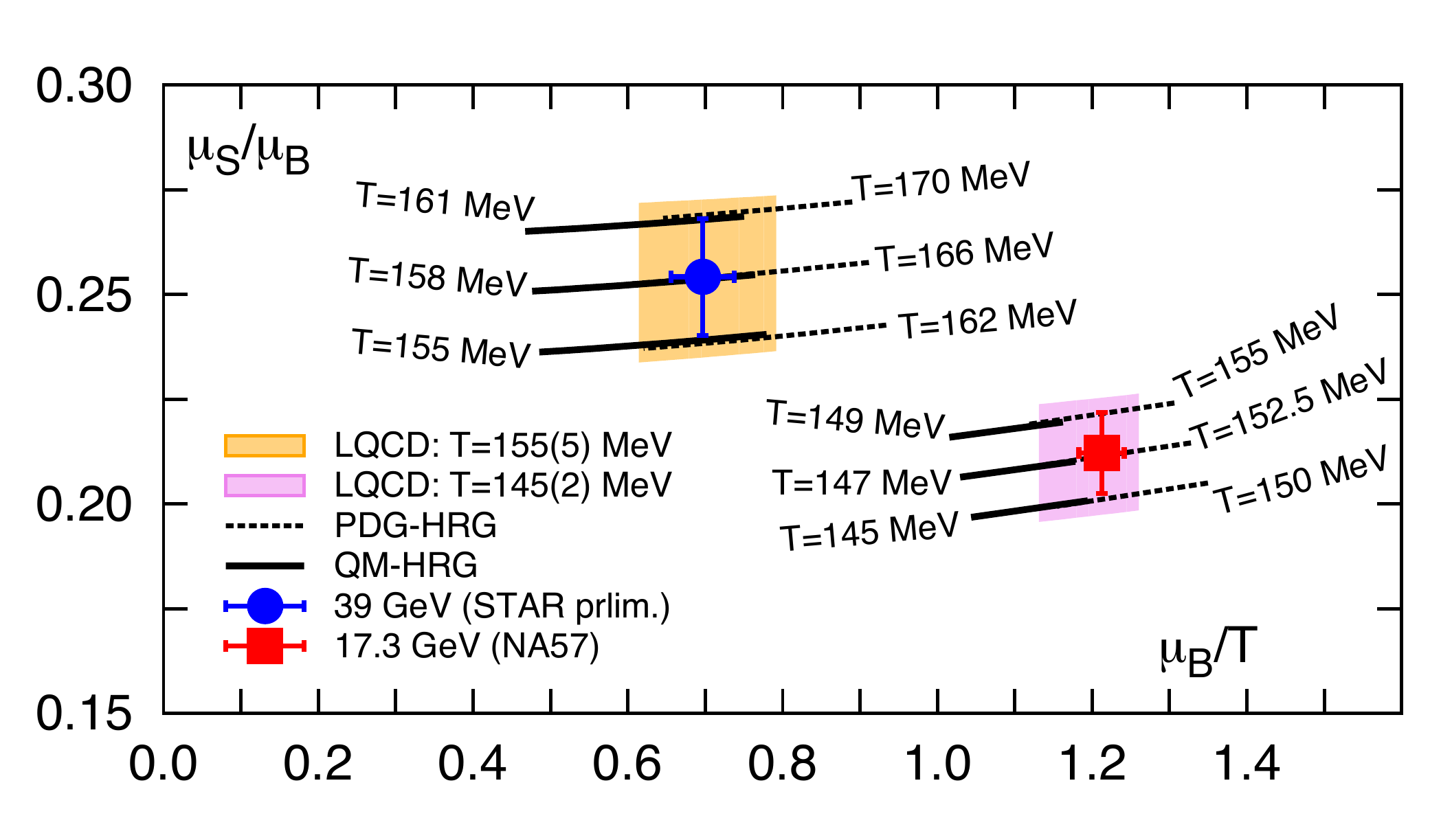}
\end{minipage}
\caption{ (Left) Lattice QCD results for $\mu_S/\mu_B$ at the leading order, {\it
i.e.} $s_1(T)$ (see Eq. [\ref{eq:muS}]). The dotted line (PDG-HRG) shows the results
of hadron resonance gas model containing only hadrons listed by the Particle Data
Group. The solid line (QM-HRG) depicts the result for a hadron gas when additional,
yet unobserved, quark model predicted strange hadrons are included
\cite{Bazavov:2014xya}. The shaded region indicate the chiral crossover region
$T_{pc}=154(9)$ MeV.  (Right) The figure shows a comparison between the experimentally
extracted values of $(\mu_S^f/\mu_B^f,\mu_B^f/T^f)$ (filled points) with the lattice
QCD results for $\mu_S/\mu_B$ (shaded bands)~\cite{Bazavov:2014xya}.  The lattice QCD
results are shown for $\mu_B/T=\mu_B^f/T^f$. The temperature range where lattice QCD
results match with $\mu_S^f/\mu_B^f$ provide the values of $T^f$, {\it i.e.}
$T=155(5)$ MeV and $145(2)$ MeV for $\sqrt{s}=39$ GeV and $17.3$ GeV, respectively.}  
\label{fig:muS_lqcd}
\end{center}
\end{figure}

It is evident from Figure~\ref{fig:fo_lqcd}, that the large errors on the
experimental values $(S_Q \sigma_Q^3)/M_Q$, allow at present, only
an upper bound on the freeze-out temperature $T^f$. 
Very recently, an alternative procedure for
the determination of $T^f$ has been demonstrated~\cite{Bazavov:2014xya}. This
procedure utilizes the fact that the initially colliding nuclei in
heavy ion collisions are free of
net strangeness and conservation of strangeness under strong interaction ensures that
the medium created during heavy ion collisions is strangeness neutral. By imposing a strangeness
neutrality condition for a homogeneous thermal medium $\left\langle n_S \right\rangle
(\mu_B,\mu_S)=0$, the strangeness chemical potential, $\mu_S$, can be determined by
performing a Taylor expansion of the net strangeness density, $\left\langle n_S
\right\rangle(\mu_B,\mu_S)$~\cite{Bazavov:2014xya} 
\begin{equation}
\frac{\mu_S}{\mu_B} = s_1(T) + s_3(T) \left(\frac{\mu_B}{T}\right)^2 +
\mathcal{O}\left[\left(\frac{\mu_B}{T}\right)^4 \right] \;.
\label{eq:muS}
\end{equation} 
Since the coefficients $s_1$, $s_3$, {\it etc.} consist of various generalized baryon,
charge and strangeness susceptibilities defined at vanishing chemical potentials, they
can also be obtained from standard lattice QCD computations at zero chemical
potentials. The leading order $s_1(T)$ coefficient for
$\mu_S/\mu_B$ is shown in Figure \ref{fig:muS_lqcd} (left).
It is interesting that comparisons of these lattice results with the
predictions from the hadron
resonance gas model reveal that the inclusion of only experimentally observed hadrons
fails to reproduce the lattice data around the crossover
region. However, the inclusion of
additional, unobserved strange hadrons predicted within the quark model provides a much
better agreement with lattice results, hinting that these additional hadrons become
thermodynamically relevant close to the crossover temperature~\cite{Bazavov:2014xya}.
Other lattice thermodynamics studies also indicate that
additional, unobserved charm hadrons also become thermodynamically relevant close to
the QCD crossover~\cite{Bazavov:2014yba}. 

In heavy ion collisions, the measured relative yields of the strangeness $S$ antibaryons to baryons
at the freeze-out are determined by the thermal freeze-out parameters
($T^f,\mu_B^f,\mu_S^f$)~\cite{Andronic:2011yq}
\begin{equation}
R_H(\sqrt{s}) = \exp\left[ -\frac{2\mu_B^f}{T^f} 
\left( 1 - \frac{\mu_S^f}{\mu_B^f}|S| \right) \right] \;.
\end{equation}
Thus, by fitting the experimentally measured values of $R_\Lambda$, $R_\Xi$ and
$R_\Omega$, corresponding to $|S|=1,2$ and $3$, at a given $\sqrt{s}$ it is easy to
determine the $\mu_S^f/\mu_B^f$ and $\mu_B^f/T^f$. By matching these experimentally
extracted values of $\mu_S^f/\mu_B^f$ with lattice QCD results for
$\mu_S/\mu_B$ as a
function of temperature, one can determine the freeze-out temperature $T^f$. Figure
\ref{fig:muS_lqcd} (right) demonstrates this procedure. Not surprisingly, the
inclusion of additional unobserved strange hadrons in the hadron resonance gas model
leads to very similar values of the freeze-out temperatures as obtained using the
lattice data. However, including only the hadrons listed in the
Particle Data Group tables~\cite{pdg:2015} yields
freeze-out temperatures that are $5-8$ MeV smaller.  

Lattice QCD calculations can also be used to locate and establish the
existence of the QCD critical point.
Early results are based
on calculations on rather small and coarse lattice using only a 1-link standard staggered 
fermion discretization scheme~\cite{Fodor:2004nz}. It has been pointed
out that the method used in this calculation, 
the determination of Lee-Yang zeroes, also suffers from an overlap
problem rather than a sign problem and may lead to spurious
signatures for a critical point~\cite{Ejiri:2005ts}. Calculations using a formulation of
finite-density QCD with an imaginary chemical potential, also performed on lattices with only
four sites in the temporal direction, do not find any evidence for
the existence of a critical point~\cite{de Forcrand:2002ci}. The most
systematic searches for a critical point at present 
are based on the Taylor series expansion of the QCD partition function 
\cite{{Gavai:2003mf},Allton:2005gk}. For vanishing electric charge and strangeness chemical
potential one can expand the pressure $P$ in terms of $(\mu_B/T)^2$, the expansion coefficients
being cumulants of net-baryon number fluctuations, i.e. generalized baryon number susceptibilities 
$\chi_n^B(T)$
\begin{equation}
\frac{P(T,\mu_B)}{T^4} = \sum_{n=0}^\infty \frac{1}{n!} \chi_n^B(T) \left( \frac{\mu_B}{T} \right)^n
\; . 
\label{mu-pressure}
\end{equation}
This series expansion has a radius of convergence which may be estimated using a finite,
typically small, set of values for the generalized baryon number susceptibilities. Subsequent estimators
\begin{equation}
r_n\equiv \left( \frac{\mu_B^{E}}{T} \right)_n = 
\sqrt{\frac{n (n-1) \chi_n^B}{\chi_{n+2}^B}} \; ,\; n~{\rm even}
\; ,
\end{equation}
may stay finite or diverge in the limit $n\rightarrow \infty$. Current estimates for 
the radius of convergence~\cite{Gavai:2008zr}, also based on calculations with the unimproved 
1-link staggered
fermion action and moderately light quark masses ($M_\pi\simeq 230$~MeV), suggest for the 
coordinates of the critical point  
$(T^E/T_{pc},\mu_B^E/T^E)= (0.94(1),1.8(1))$.  However, true
systematic errors for these calculations are difficult to estimate,
and this topic is currently under active investigation.

%
%
\section{TRANSPORT PROPERTIES AND HEAVY QUARKS}
\label{sec:summary}

This review focuses on recent results on bulk QCD thermodynamics with physical quark
masses and on fluctuations of conserved charges. These two topics are of immediate
importance for the ongoing experimental studies on the phase structure of
strong-interaction matter. However, a comprehensive understanding of the strongly
coupled nature of QGP for temperatures $T_{pc}\lesssim T \lesssim 2T_{pc}$
requires lattice QCD calculations of its color screening and transport
properties.  Many of these calculations are performed in the quenched
approximation, in which the dynamical fermion loops are neglected.
Furthermore, 
full continuum extrapolations have not yet been performed, but they
are often performed on large lattices that are close enough to the continuum
to yield meaningful results. 
We give a brief summary of these important
calculations and the insights that they provide.

\subsection{Color Screening}

Matsui and Satz pointed out that the force between heavy quarks inside
a QGP is
screened due to presence of color charges and eventually leads to the dissolution of
quarkonia, bound states of heavy quark and anti-quark, such as $J/\psi$, $\eta_c$,
$\Upsilon$  {\it etc.}~\cite{Matsui:1986dk}. Lattice QCD calculations of the
potential between two infinitely heavy static quarks have established this color
screening mechanism~\cite{Bazavov:2013zha,Burnier:2014ssa}.  However, quantitative
understanding of the dissociation temperatures of quarkonia require knowledge of the
spectral functions of quarkonia. 
Extraction of these real (Minkowski) time quantities from the Euclidean time quarkonia
correlation functions measured on the lattice require analytic continuations to
Minkowski time. Since the number of lattice points along the Euclidean time direction
are limited, such analytic continuations are usually performed using a Bayesian
method, such as the Maximum Entropy Method~\cite{Asakawa:2000tr,Rothkopf:2011ef}.
Reliable analytic continuation via the Maximum Entropy Method demands lattice data at
large numbers of Euclidean time points, and, hence, lattices with large temporal
extents. Calculations of charmonium spectral functions in quenched QCD
have been performed on large
lattices, close to the continuum limit~\cite{Ding:2012sp}. These calculations suggest
that both P- and S-wave ground state charmonia disappear in a QGP at temperatures
$T\gtrsim 1.5 T_{pc}$. Calculations with $2+1$ flavors of dynamical quarks, but with
un-physically heavy pion masses, have started
\cite{Skullerud:2014sla,Borsanyi:2014vka} and lead to similar conclusions. Lattice
QCD study of the spatial correlation functions of charmonia with $2+1$ dynamical
flavors having almost physical quark masses has also provided indirect evidence that
the ground state charmonia cease to exist inside QGP for $T\gtrsim1.3T_{pc}$
\cite{Bazavov:2014cta}. At present, lattice spacings ($a$) used in the
state-of-the-art lattice QCD calculations at non-zero temperatures are
still too large such that the bottom quark mass ($m_b$) in lattice units are $am_b\gtrsim1$.
This leads to large cut-off effects for studies related to bottomonia spectral
functions. Thus, currently one uses a hybrid approach where the heavy bottom quarks
are treated within the non-relativistic QCD (NRQCD) approximation.  However,
the NRQCD formalism does not possess a proper continuum limit. These
calculations~\cite{Skullerud:2014sla,Aarts:2014cda,Kim:2014iga} suggest that the
ground state S-wave bottomonium survive up to $\sim2T_{pc}$, but
the implications for the P-wave states are still unclear.

\subsection{Transport Coefficients}

Calculations of transport coefficients of QGP using Euclidean time lattice QCD also
require analytic continuation to real Minkowski time, and therefore
also depend on the extraction of spectral functions. Shear and bulk viscosity can be obtained from
Euclidean time correlation functions of the energy-momentum tensor. Since the
energy-momentum tensor operator is dominated by purely gluonic correlation functions
these correlation functions tend to be very noisy and calculations of shear and bulk
viscosities on the lattice remain extremely challenging. Although, there were
attempts to calculate shear and bulk viscosities for a pure gauge
theory~\cite{Karsch:1986cq,Nakamura:2004sy,Meyer:2007ic,Meyer:2007dy},
so far,  there is no lattice calculation of these quantities for realistic QCD. 

Transport coefficients
determined solely by quark operators, such as the electrical conductivity, are more
accessible from present day lattice QCD. Calculations of electrical conductivity
($\sigma$) within the quenched approximation are quite advanced and results close to
the continuum limit exist~\cite{Ding:2010ga}.  A recent calculation
\cite{Ding:2014dua} at three temperatures in the range $T= 1.1T_{pc}-1.5T_{pc}$ constrains
the value to a rather narrow range, $0.2 C_{em} \lesssim \sigma/T \lesssim 0.4
C_{em}$, where $C_{em}$ denotes the sum over the squared electric charges of quarks.
More realistic calculations of $\sigma$ with dynamical fermions have also started to
become available in recent years~\cite{Brandt:2012jc,Amato:2013naa,Aarts:2014nba} and
show a striking drop in the value of $\sigma/T$ when a approaching $T_{pc}$ from high
temperatures~\cite{Amato:2013naa,Aarts:2014nba}. Similar calculations also provide
the charge diffusion constant~\cite{Aarts:2014nba} and the thermal di-lepton
production rate in QGP~\cite{Ding:2010ga,Ding:2014dua}.

Spectral functions associated with the vector current for the charm
quark also provide access to the charm quark diffusion constant
($D$)~\cite{Ding:2012sp}.  The momentum 
diffusion coefficient of an infinitely heavy quark, $\kappa = 2T^2/D$, can also be
extracted from the correlation function of purely gluonic operators under the heavy
quark effective theory approximation~\cite{CaronHuot:2009uh}. Current results
\cite{Banerjee:2011ra,Kaczmarek:2014jga} on the momentum diffusion constant of
infinitely heavy quark are consistent, yielding $\kappa \simeq 2.5 T^3$. The
corresponding diffusion constant $D\simeq 0.8/T$ is about a factor two larger than
the charm quark diffusion constant extracted using the charm vector current
correlation function~\cite{Ding:2012sp}.

The jet quenching parameter $\hat{q}$, an important ingredient in the analysis of
energy loss of jets traversing QGP, may also become accessible to lattice QCD
calculations. Calculation of this quantity involves correlation functions of Wilson
lines along the light-cone and extraction of that from Euclidean lattice QCD again
demands analytic continuation to real time. One way of performing such analytic
continuation can be justified at very high temperature where the weak coupling
expansion is valid~\cite{CaronHuot:2008ni}. Following this proposal, a part of
the non-perturbative contributions to the $\hat{q}$ was calculated recently within
the dimensionally reduced effective theory, electrostatic QCD~\cite{Panero:2013pla}.
If this proposal can be extended to full QCD and down to the truly non-perturbative
regime close to the QCD crossover then it will open up a new avenue for lattice QCD
calculations that will directly impact the phenomenology of strongly interacting
matter probed in heavy ion collisions.
\section{CONCLUSIONS AND OUTLOOK}
\label{sec:conclusion}

Calculations of the fundamental thermodynamic quantities of QCD as
presented in this review have reached a significant milestone.  The
crossover temperature and equation of state have been calculated with
physical values for the light and strange quark masses and separate, reliable
continuum extrapolations have been performed by two collaborations.
The transition is firmly established as a crossover, as evident in
both the equation of state results and observation that chiral
susceptibility is independent of volume.  The latter result has been
achieved with fermions that differ in their approach to chiral
symmetry: a staggered fermion action in which full chiral symmetry is
restored in the continuum, and the domain wall action in which
chiral symmetry is preserved for finite lattice spacings.
Because the transition is a crossover, the definition of a
transition temperature is quantity dependent.  However,
the chiral condensate, which is the order parameter for the phase
transition in the chiral limit, is a natural choice.  For this
quantity there is also remarkable agreement among the recent
calculations despite significant differences in the analysis methods
used.  A fit to the inflection point in the renormalized chiral
condensate with the stout action yielded $T_{pc}=155 \pm 2 \pm 3$~MeV.  A
more sophisticated analysis involving scaling fits to the $O(N)$
universality class for a constrained continuum extrapolation to the
HISQ and asqtad actions produced a value of $T_{pc}=154 \pm 8 \pm 1$~MeV.
The general range for $T_{pc}$ has also been reproduced in a domain
wall calculation with physical quark masses and a lattice spacing
$N_\tau=8$.

The close agreement  between different actions and analysis methods extends to the
equation of state, where continuum extrapolations with the stout and
HISQ actions agree to within their respective errors over
the temperature range 130--400~MeV.  Although a small difference
begins to develop at the higher end of this temperature range, it is
not yet known whether this will lead to a more significant
difference above this temperature range.  At this time the overall
precision and estimated accuracy of the crossover temperature and
equation of state at zero baryon density appear to be sufficient to
meet the needs of the heavy ion community.  The need for future
improvements from the lattice will depend upon the sophistication of
the phenomenological tools and the desired accuracy for extracted
physics parameters. 

Calculations of fluctuations on the lattice are a more recent
development, and this area has received considerable attention and
resources only within the past few years.  The ability to calculate
freeze-out curves and net-charge moments that can be directly compared
with heavy ion experiments represents a significant advance, and one
the will hopefully elucidate the location and signatures of the
critical point.  At this time predictions of the critical point in the
$T$-$\mu$ plane are highly uncertain, and significant advances are
required in both computing power and algorithm efficiency if this goal
is to be attained within the next several years.

Finally, calculations of light and heavy quark bound states,
diffusion, and now jet quenching hold considerable promise for the
future, and one can expect significant results to follow when full QCD
calculations with physical quark masses are possible. 

This work is supported in part through contracts No. DE-AC52-07NA27344
and No. DE-SC0012704 with the U.S. Department of Energy
and NSF Grant No. PHY10-034278.



\end{document}